\documentclass[]{aa}

\usepackage{graphicx}
\usepackage{txfonts}
\usepackage[colorlinks=true,allcolors=blue]{hyperref}

\usepackage{natbib}
\usepackage{multirow}
\usepackage{multicol}
\usepackage{textcomp}
\usepackage{xcolor}
\usepackage{verbatim}
\usepackage[labelfont=bf]{caption}
\usepackage{longtable, booktabs}
\usepackage{lscape}
\usepackage{enumitem} 
\makeatletter
\renewcommand*\aa@pageof{, page \thepage{} of \pageref*{LastPage}}
\makeatother

\def\Msun{M_\odot}
\def\MJ{M_\mathrm{J}}
\def\Mp{M_\mathrm{p}}
\def\Macc{\dot{M}_\mathrm{acc}}
\def\RJ{R_\mathrm{J}}
\def\Rp{R_\mathrm{p}}
\def\Teff{T_{\text{eff}}}
\def\Rd{R_\mathrm{dust}}
\def\Rg{R_\mathrm{gas}}
\def\micron{\mu\mathrm{m}}
\def\Lacc{L_\mathrm{acc}}

\DeclareUnicodeCharacter{2212}{-}
\begin{document}

\title{ISPY- NACO Imaging Survey for Planets around Young stars\\ The demographics of forming planets embedded in protoplanetary disks \thanks{Based on observations collected at the Paranal Observatory, ESO (Chile). Program ID: 097.C-0206(A), 097.C-0206(B), 198.C-0612(A), 198.C-0612(B), 198.C-0612(C), 199.C-0065(A), 199.C-0065(A2), 199.C-0065(B), 199.C-0065(C), 199.C-0065(D), 1101.C-0092(A), 1101.C-0092(C), 1101.C-0092(D), 1101.C-0092(E), 1101.C-0092(F), 1101.C-0092(G), 1101.C-0092(H).}\thanks{The reduced images and contrast curves are only available at the CDS via anonymous ftp to cdsarc.u-strasbg.fr (130.79.128.5) or via https://cdsarc.u-strasbg.fr/viz-bin/qcat?J/A+A/}}

\author{G. Cugno\inst{\ref{ethz}, \ref{Umich}}
\and T. D. Pearce\inst{\ref{jena}} 
\and R. Launhardt\inst{\ref{mpia}}
\and M.~J.~Bonse\inst{\ref{ethz}} 
\and J. Ma\inst{\ref{ethz}}
\and T. Henning\inst{\ref{mpia}}
\and A. Quirrenbach\inst{\ref{LSW}}
\and D. S\'egransan\inst{\ref{unige}}
\and E.~C.~Matthews\inst{\ref{unige}}
\and S.~P.~Quanz\inst{\ref{ethz}}
\and G.~M.~Kennedy\inst{\ref{warwick1},\ref{warwick2}}
\and A. M\"uller\inst{\ref{mpia}}
\and S. Reffert\inst{\ref{LSW}} 
\and E.~L.~Rickman\inst{\ref{stsci}}
}

\institute{ETH Zurich, Institute for Particle Physics and Astrophysics, Wolfgang-Pauli-Strasse 27, CH-8093 Zurich, Switzerland\label{ethz}
\and Department of Astronomy, University of Michigan, Ann Arbor, MI 48109, USA \label{Umich}
\and Astrophysikalisches Institut und Universit\"atssternwarte, Friedrich-Schiller-Universit\"at Jena, Schillerg\"sschen 2–3, 07745 Jena, Germany \label{jena}
\and Max-Planck-Institut f\"ur Astronomie, K\"{o}nigstuhl 17, 69117 Heidelberg, Germany \label{mpia}
\and Landessternwarte, Zentrum f\"ur Astronomie der Universit\"at Heidelberg, K\"onigstuhl 12, 69117 Heidelberg, Germany\label{LSW}
\and Observatoire Astronomique de l’Universit\'e de Gen\`eve, 51 Ch. des Maillettes, 1290 Versoix, Switzerland\label{unige}
\and Department of Physics, University of Warwick, Gibbet Hill Road, Coventry CV4 7AL, UK\label{warwick1}
\and Centre for Exoplanets and Habitability, University of Warwick, Gibbet Hill Road, Coventry CV4 7AL, UK\label{warwick2}
\and European Space Agency (ESA), ESA Office, Space Telescope Science Institute, 3700 San Martin Drive, Baltimore, MD 21218, USA
\label{stsci}
\\\\\email{gcugno@umich.edu}
}

%

\date{Received --- ; accepted --- }

\abstract{}{}{}{}{}
\abstract
{Planet formation is a frequent process, but little observational constraints exist about the mechanisms involved, especially for giant planets at large separation. The NaCo-ISPY large program is a 120 night $L'$-band direct imaging survey aimed at investigating the giant planet population on wide orbits ($a>10$~au) around stars hosting disks.}
{Here we present the statistical analysis of a subsample of 45 young stars surrounded by protoplanetary disks (PPDs). This is the largest imaging survey uniquely focused on PPDs to date. Our goal is to search for young forming companions embedded in the disk material and to constrain their occurrence rate in relation to the formation mechanism.}
{We used principal component analysis based point spread function subtraction techniques to reveal young companions forming in the disks. We calculated detection limits for our datasets and adopted a black-body model to derive temperature upper limits of potential forming planets. We then used Monte Carlo simulations to constrain the population of forming gas giant companions and compare our results to different types of formation scenarios. }
{Our data revealed a new binary system (HD38120) and a recently identified triple system with a brown dwarf companion orbiting a binary system (HD101412), in addition to 12 known companions. Furthermore, we detected signals from 17 disks, two of which (HD72106 and T\,CrA) were imaged for the first time. We reached median detection limits of $L'=15.4$ mag at $2\farcs0$, which were used to investigate the temperature of potentially embedded forming companions. We can constrain the occurrence of forming planets with semi-major axis $a$ in $[20-500]$~au and $\Teff$ in $[600-3000]$~K to be $21.2^{+24.3}_{-13.6}$\%, $14.8^{+17.5}_{-9.6}$\%, and $10.8^{+12.6}_{-7.0}$\% for $\Rp = 2,3,5~\RJ$, which is in line with the statistical results obtained for more evolved systems from other direct imaging surveys. These values are obtained under the assumption that extinction from circumstellar and circumplanetary material does not affect the companion signal, but we show the potential impact these factors might have on the detectability of forming objects. }
{The NaCo-ISPY data confirm that massive bright planets accreting at high rates are rare. More powerful instruments with better sensitivity in the near- to mid-infrared (MIR) are likely required to unveil the wealth of forming planets sculpting the observed disk substructures. }

\keywords{Techniques: high angular resolution -- Planets and satellites: detection, formation}

\titlerunning{The demographics of forming planets embedded in protoplanetary disks}
\maketitle
\definecolor{green2}{rgb}{0,0.8,0.2}

\section{Introduction}
\label{Introduction}

In the last two and a half decades, more than 5000 extrasolar planets have been discovered. Their detection revealed a breathtaking diversity in their characteristics, such as mass, orbital separation, density, and atmospheric properties \citep[e.g.,][]{Winn2015, Kaltenegger2017, Madhusudhan2019}. Furthermore, we now know that planet formation is a very frequent and efficient process. Most of the known exoplanets have been discovered with transit and radial velocity (RV) surveys, and in the future many more will be revealed thanks to ongoing and future missions such as {\it Gaia}, which is expected to significantly contribute to the exoplanet inventory  \citep[e.g.,][]{Perryman2014, Arenou2022}. These methods provide information on the planet demographics, but they are not able to put direct empirical constraints on their formation, as they suffer from observational biases that make observing young stars not ideal.

Complementary to the RV, astrometry, and transit techniques, direct imaging prefers young (i.e., brighter) planets well separated ($a>10$~au) from their host star. In recent years, huge efforts and resources have been deployed to improve essential steps such as adaptive optics (AO) systems \citep[e.g.,][]{Beuzit2019}, coronagraphy \citep[e.g.,][]{Martinache2019}, and post-processing analysis \citep[e.g.,][]{Cantalloube2021}, which are indispensable to reach the high contrast necessary to image forming planets. 

All of these investments brought several discoveries, including the iconic $\beta$ Pic system \citep{Lagrange2009} and the HR8799 system, where at least four giant planets are orbiting the same star \citep{Marois2008, Marois2010, Wang2018}. Focusing on forming giant planets, several candidates were proposed in the past decade, but unfortunately most of them remain unconfirmed and under debate \citep[e.g., LkCa15\,b, HD100546\,b and c, and HD169142\,b;][to name a few]{Sallum2015, Quanz2013, Reggiani2014, Biller2014, Rameau2017, Reggiani2018}. The first confirmed imaged forming planets have been PDS70\,b \citep{Keppler2018, Muller2018} and PDS70\,c \citep{Haffert2019}. Since then, multiple studies to unveil their nature have been conducted \citep[e.g.,][]{Bae2019, Stolker2020_pds70, Wang2021, Cugno2021, Benisty2021}. Finally, more recently, another protoplanet has been proposed around AB Aur using observations from multiple instruments over several years \citep{Currie2022_ABAur}. 

Direct imaging young, still forming planets is a particularly difficult, and yet important, task. Indeed, being embedded in the disk material, the planet flux is expected to suffer from extinction. Furthermore, high-contrast imaging post-processing techniques, especially angular differential imaging \cite[ADI;][]{Marois2006}, might distort the disk scattered light morphology making it appear as point sources in the final residual images \citep[e.g.,][]{Follette2017, Ligi2018}. On the positive side, thermal emission from circumplanetary material is expected to contribute at longer wavelengths, potentially enhancing the possibility of a discovery \citep[e.g.,][]{Zhu2015}. 
Despite these obstacles, detecting and studying forming planets will shed light on what are the main mechanisms that are driving planet formation, which formation model is expected to dominate \citep[gravitational instability vs. core accretion;][]{Boss1997, Pollack1996}, whether planets form following a hot-start, cold-start, or warm-start scenario \citep{Marley2007, Spiegel2012, Mordasini2012}, and where planets form with respect to their host star, potentially as a function of host star properties. 

These are some of the questions that motivated the NaCo Imaging Survey for Planets around Young stars \citep[NaCo-ISPY;][]{Launhardt2020}, an $L'$ imaging campaign at the Very Large Telescope (VLT) in Chile where we used 120 nights to investigate the population of gas giant planets around disk-hosting stars. The ISPY targets can be divided into two different classes, debris disks (DEBs, 203 targets) and protoplanetary disks (PPDs, 50 targets).

While several other (larger) surveys tried and are trying to provide statistical constraints on the overall population of giant planets \citep[e.g.,][]{Chauvin2010, Brandt2014, Chauvin2015, Stone2018, Nielsen2019, Vigan2021}, NaCo-ISPY is unique in the sense that it only focuses on stars hosting a disk, trying to exploit this particularity while studying the potential interaction between disks and planets \citep{MussoBarcucci2019, Pearce2022}. A smaller survey with 15 targets and similar aims has been recently conducted with the VLT/SPHERE instrument by \cite{Asensio-Torres2021}, unfortunately without detecting new planets. In Sect.~\ref{sec:targets} we present the sample of the survey, and in Sect.~\ref{sec:observations} we describe the observations, with the data reduction detailed in Sect.~\ref{sec:data_reduction}. Section~\ref{sec:individual_objects} discusses results for individual targets, while Sect.~\ref{sec:results_all} presents the statistical results of the survey as a whole. These results are discussed in Sect.~\ref{sec:discussion} and we present our conclusions in Sect.~\ref{sec:conclusions}.

\section{Survey sample}
\label{sec:targets}
\subsection{Initial target list}
The process of the target selection for the NaCo-ISPY survey has been detailed in \cite{Launhardt2020}. As in this paper we only focus on the PPDs targets, we here provide some details on this subsample, and we refer to \cite{Launhardt2020} and \cite{Pearce2022} for information on the debris disk sample. In short, we compiled the initial target list from studies of Herbig Ae/Be stars \citep{The1994, Menu2015}, motivated by early detections of protoplanetary candidates orbiting those type of stars \citep[e.g.,][]{Kraus2012, Quanz2013, Reggiani2014}. These lists were complemented with additional objects hosting structured disks, which could indicate ongoing planet formation \citep[e.g.,][]{ALMA2015, vanBoekel2017, Konishi2016, Andrews2018, Avenhaus2018}. 
Additional requirements were the target declination (-70$^\circ<$DEC$<+15^\circ$), distance ($d<1000$ pc as measured at the time of compilation), multiplicity and $K$-band magnitude to ensure high quality AO correction \citep{Launhardt2020}. 
A total of 90 targets hosting a PPD were identified. We discarded 14 objects from our list because of existing $L'$ data or limited discovery space. Out of the 76 remaining targets, during the survey we observed 50 objects.

\subsection{Sample properties}
Targets of our sample are listed in Table \ref{tab:target_list} and their parameter distributions are shown in Fig.~\ref{fig:system_properties}. In the next paragraphs, we provide a top-level overview of these parameters, their source and their relevance for the ISPY search for forming planets.

Distances (top left panel of Fig.~\ref{fig:system_properties}) were obtained from {\it Gaia} DR3 \citep{Gaia2022} unless stated otherwise. Most of our targets have distance $d<200$ pc, but some are much farther away.

$L'$ magnitudes (top right panel of Fig.~\ref{fig:system_properties}) are obtained interpolating WISE photometry \citep{Cutri2013} between the W1 (3.35~$\micron$) and W2 (4.6~$\micron$) filters to the wavelength of our observations (3.8~$\micron$). The vertical line in the $L'$ histogram represents the limit for the use of the Annular Groove Phase Mask (AGPM) coronagraph: for stars brighter than $L'\sim6.5$~mag we tried to use the coronagraph to increase contrast performance close to the star (see Sect.~\ref{sec:observations} for details about the observational setup). We note, however, that the final choice on the use of the coronagraph also depended on the weather conditions during the data acquisition.

Spectral types are taken from the SIMBAD database \citep{Wenger2000}, and span a wide range, from M3 to B5. Correspondingly, effective temperatures range from 3900~K to 14,000~K (middle left panel of Fig.~\ref{fig:system_properties}, \citealt{Launhardt2020, Pearce2022}).  

Stellar masses (middle right panel of Fig.~\ref{fig:system_properties}) are taken from \cite{Kervella2019}, but we used other literature values whenever unavailable. From other large direct imaging and ALMA surveys we know that massive stars have a higher chance to host giant planets \citep[e.g.,][]{Vigan2021, Janson2021, Squicciarini2022} and massive disks \citep{Andrews2013, Ansdell2016}, which provide the building blocks to form planetary systems. Thus, it is a natural choice to focus on massive young Herbig Ae/Be stars in a survey such as NaCo-ISPY, and indeed almost all of our targets have masses $M_*\gtrsim 1 M_\odot$. 

Stellar age is a crucial parameter for the interpretation of detection limits in high-contrast imaging data, as the magnitude to mass conversion strongly depends on the age assumption, especially for young targets. At the same time, age is one of the most difficult stellar parameters to constrain for young objects, with uncertainties and biases depending on the applied methods that dominate the measurements. We compiled stellar ages from the literature (see Table \ref{tab:target_list}), even though in Sect.~\ref{sec:results_all} we constrain the population of forming protoplanets independently from age estimates.

\subsection{Disks}
\label{sec:disks}
A large fraction of our targets has been imaged with high resolution observations tracing the mm dust in thermal continuum, the scattering dust or the disk gas phase.
Substructures in those components have been found to be ubiquitous. For example, rings, gaps and cavities are found in almost every disk imaged at sufficiently high resolution \citep[e.g.,][]{Garufi2018, Andrews2018, Law2021_MAPS3}. Hydrodynamical simulations \citep[e.g.,][]{Zhu2012, Dipierro2016} indicate that they could be the result of protoplanets sculpting the disk material: indeed, young companions create gas pressure bumps able to stop dust radial drift and thus trap the dust in ring-like structures \citep[e.g.,][]{Pinilla2015, Bae2018}. Other commonly detected structures are spiral arms, which have been detected in several disks \citep{Muto2012, Isella2018, Teague2019, Muro-Arena2020}. Similarly to gaps, one of the most convincing explanations for their existence is the interaction with embedded planets \citep{Fung2015, Bae2016}. Additionally, shadows have been observed in several disks \citep{Stolker2017, Bohn2022, Teague2022}, which could be caused by the presence of a warped unresolved inner disk that underwent dynamical interaction with a forming planet \citep{Nealon2018}. Despite the presence of forming planets being a very exciting explanation for all the disk substructures observed in the last decade, other mechanisms could be able to explain the disk observations and should be considered \citep[e.g.,][]{Zhang2015, Birnstiel2015, Paneque-Carreno2021}.

\begin{figure}[t!]
    \includegraphics[width=\hsize]{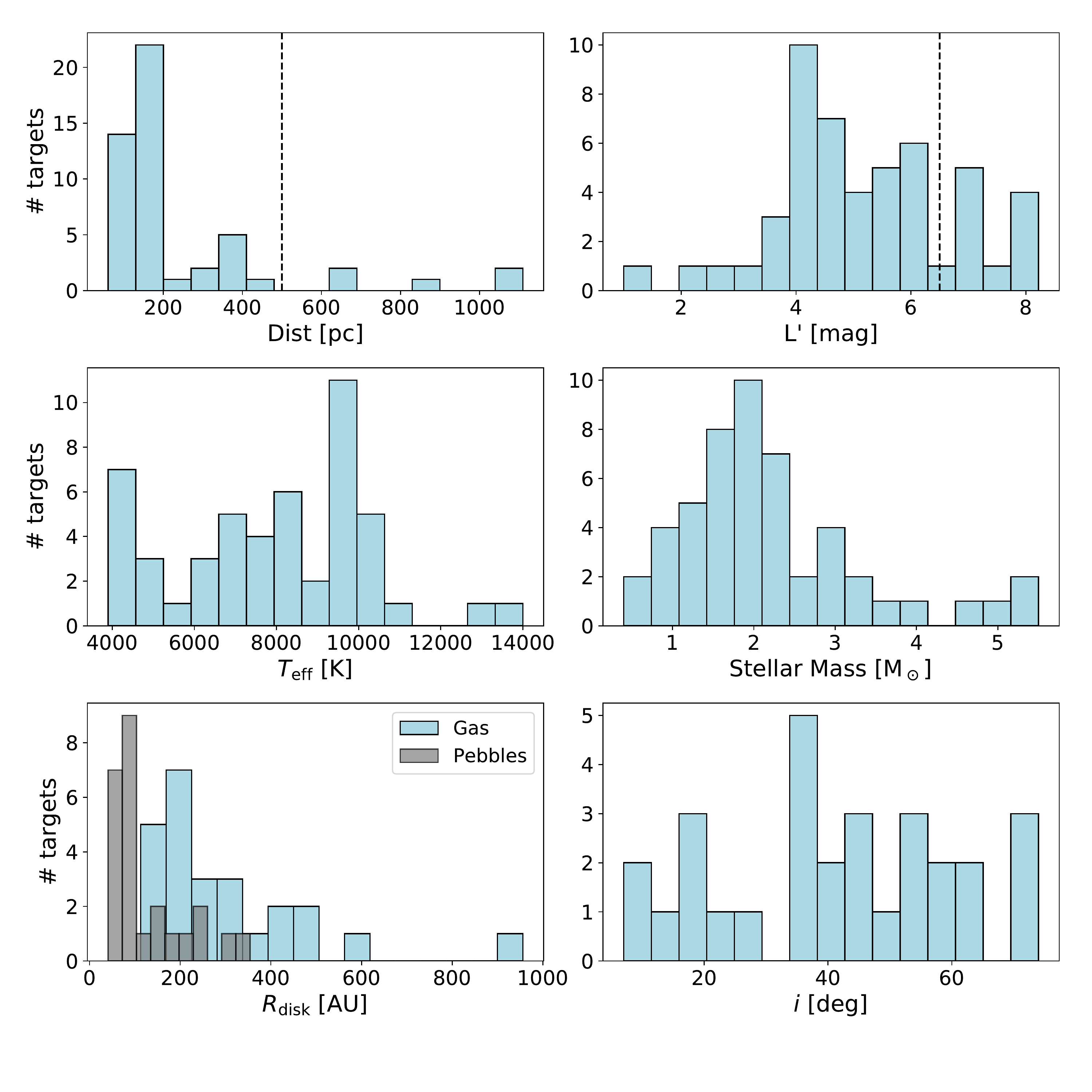}
    \caption{Stellar and disk parameters for the objects in the ISPY PPD sample. In the top row the histogram for the distances (left) and $L'$ observed magnitudes (right) are reported. Stellar effective temperatures (left) and masses (right) are shown in the middle row. Finally, the bottom row provides the distributions of the disk outer radius (gas and mm-dust) on the left and disk inclinations (whenever measured) on the right. The dashed vertical lines represent the distance cutoff applied in Sect.~\ref{sec:additional_selection} ({\it top left panel}) and the limiting brightness for the coronagraphic observations ({\it top right panel}), respectively.}
    \label{fig:system_properties}
\end{figure}

Given our goal of investigating the formation of planets in their natal protoplanetary disks, our search space is limited by the disk outer edge (see Table~\ref{tab:target_list}). We used disk radii found in the literature from high angular resolution data obtained either with high contrast imagers or with ALMA and corrected for the newly measured distance from {\it Gaia} DR3. We confined our search region to a circular aperture from the central star with radius equal $1.5\times\Rg$, where the factor $1.5$ has been conservatively included to ensure that we are tracing the full radial extent of the disk and we do not reduce our search space because of low sensitivity observations. For some disks, measurements of $\Rg$ do not exist, and we extrapolated the disk radius from measurements of the dust disk radius $\Rd$ traced by the (sub)millimeter continuum emission. 
Several observational and theoretical studies indicate that gas in protoplanetary disks has a much larger extent than pebbles \citep[e.g.,][]{Andrews2012, Isella2012, Birnstiel2014, Cleeves2016, Law2021_MAPS3, Zormpas2022}. According to \cite{Trapman2019}, the dichotomy in dust and gas sizes is due to a difference in optical depth between the two components \citep[e.g.,][who estimated $\Rg/\Rd$ between $\sim1.4$ and $\sim4$ depending on the disk turbulence]{Facchini2017} and grain growth and subsequent radial drift \citep[e.g.,][]{Natta2004, Ricci2010}. \cite{Ansdell2018} investigated the relationship between sizes of the dust and gas components. They found gas-to-dust size ratios $\Rg/\Rd$ between 1.5 and 3.5, with an average of $\langle\Rg/\Rd\rangle=1.96\pm0.04$. Similarly, \cite{Long2022} estimated a $\langle\Rg/\Rd\rangle$ ratio of $2.9\pm1.2$ for 44 protoplanetary disks around stars with masses of $0.15-2.0~\Msun$ and ages of $0.5-20$~Myr. \cite{Trapman2019} estimated $\Rg/\Rd$  using analytical models, and found that the ratio has a value usually between 1.5 and 3.5, if no dust evolution has already occurred. Based on these results, when only mm continuum measurements were available, we conservatively assumed a ratio of $\Rg/\Rd=3.5$. 
A total of 20 targets (see Table~\ref{tab:target_list}) do not have disk size measurements at all; in those cases we used the median of the measured disk sizes $\overline{\Rg}=240$~au multiplied by 1.5. The distribution of disk radii for our targets is shown in the bottom left panel of Fig.~\ref{fig:system_properties} for the pebbles and the gas/$\mu$m-sized dust. As expected, the distributions indicate that overall $\Rd<\Rg$.

Disk inclinations are compiled from the literature (see references in Table~\ref{tab:target_list} and the bottom right panel of Fig.~\ref{fig:system_properties}) with values obtained from spatially resolved high angular resolution imaging data. For two targets we imaged for the first time the disk ring in scattered light (HD72106, T\,CrA), see Sect.~\ref{sec:disk_det}. For those targets we estimated the disk inclinations, which are reported in Table~\ref{tab:target_list}. Despite  not being one of the main parameters used for target selection, disk inclination plays an important role in the detectability of planetary systems if one assumes that the orbital plane is coplanar with the protoplanetary disk. First, a highly inclined disk is very optically thick, preventing the light of the planet to escape. Second, a planet in a face-on disk is potentially always visible if it is above the contrast limits, while in an inclined disk, it will spend a considerable fraction of its orbit at smaller projected angular separations from the central star, in regions not accessible by the observations due to higher contrast and angular resolution requirements. 

\subsection{Additional target selection}
\label{sec:additional_selection}
To image and investigate the early phases of giant planet formation, we need access to a considerable fraction of the region occupied by protoplanetary disks. Too large stellar distances prevent us from inspecting the inner region of infant planetary systems because of the lack of spatial resolution. Furthermore, because the flux scales with $1/d^2$, the chances of a planetary mass companion to be detected are dramatically reduced for distant targets.
Thus, we excluded from our sample 5 targets, as they have $d>500$~pc (see Table~\ref{tab:target_list}). Those targets, which are HD85567, HD259431, HD95881, HD98922 and HD190073, were initially included in the sample because at the time of the compilation of the target list precise parallax measurements from {\it Gaia} were unavailable and they were thought to be much closer. As an example, the distance of HD95881 was thought to be $170\pm30$~pc \citep{Verhoeff2010}, while {\it Gaia} locates the star at $\sim1110$~pc. 
This additional cutoff leaves us with 45 targets analyzed in this paper. However, before discarding these targets we verified that no companion candidate nor disk signal were detected in the data following the procedure described in Sect.~\ref{sec:data_reduction}.

\section{Observations}
\label{sec:observations}
Data presented in this paper were taken between 2016 May 02 and 2019 May 25 using the AO-assisted NaCo imager \citep{Rousset2003, Lenzen2003} with the $L'$ ($\lambda = 3.8\,\mu$m) filter at the Very Large Telescope (VLT) at Paranal Observatory in Chile. 
Observations were carried out in Visitor Mode, and therefore some suffered from variable or bad weather conditions. In some cases, we opted for reobservation of the same system under better conditions, and here we only present the best available dataset for each target. Table~\ref{tab:observations} reports the observations presented in this paper together with weather conditions.

All observations were taken in pupil-tracking mode to enable ADI, with field rotations depending on the elevation of the target and the integration time. We always tried to maximize the amount of rotation in the data in order to minimize self-subtraction effects when applying PSF-subtraction algorithms. For bright targets ($L'<6.5$ mag), we took advantage of the annular groove phase mask (AGPM) vector vortex coronagraph \citep{Mawet2013} to further suppress stellar diffraction at small separations and improve the contrast. During these observations, the thermal background is sampled every 13 exposures by offsetting the sky position. For fainter targets ($L'>6.5$ mag), the AGPM could not be used as the star could not be properly centered behind the mask and we dithered the star for every cube (a cube is a collection of $\sim100-120$ frames, depending on the dataset) on the three working quadrants of the L27 camera (pixel scale 0.027 mas/pix)\footnote{The bottom left quadrant suffers from bad columns and low sensitivity.}. This sequence allows us to measure the background in the quadrant with the star before and after each exposure, which can be used to reconstruct the thermal contribution in each image. 

At the beginning and at the end of each observing sequence, we took frames of the star without the coronagraph and with shorter exposure time to avoid saturation. Those frames are used to flux-calibrate each dataset. Thus, we made sure that unsaturated PSF images were particularly stable and, when necessary, we applied a strict manual selection to guarantee that variable weather conditions did not bias our results (see last column of Table~\ref{tab:observations}).

\begin{figure}[t!]
    \centering
    \includegraphics[width=\hsize]{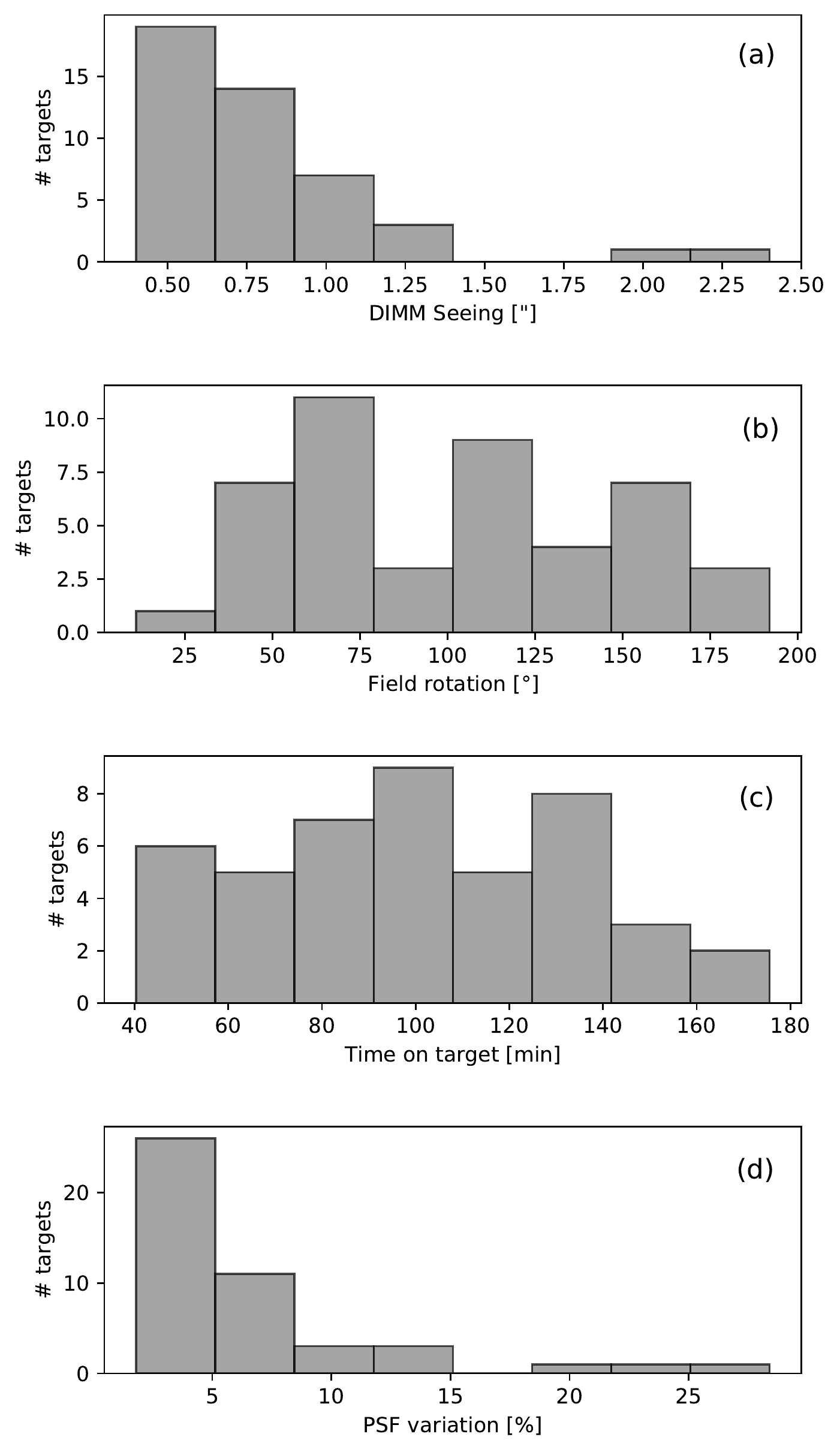}
    \caption{Histograms reporting the weather conditions at the time of observations and key parameters describing the datasets. Panel {\it (a)} reports the DIMM seeing, panel {\it (b)} the field rotation, panel {\it (c)} the time spent on each target and panel {\it (d)} the PSF variation, respectively.}
    \label{fig:obs_conditions}
\end{figure}

Figure~\ref{fig:obs_conditions} reports the most relevant properties of our observations. Overall, the median seeing was $<1\farcs2$ (panel {\it (a)}), with only two exceptions: V892~Tau and HD100453 (see also Table~\ref{tab:observations}). Also, we note that for 19 targets, observations were executed with a median seeing $\lesssim0\farcs6$. The standard deviation of the seeing during each observation is reported in Table~\ref{tab:observations} as well and can be interpreted as a measure for the stability of the atmosphere during the observing sequence. Given the relatively deep and long observations, and the fact that in most of the cases we planned observations in a period of the year with high sky rotation rate in the region of the target, we often achieved field rotations $>60^\circ$ (panel {\it (b)}). The third panel shows the time on target (ToT) we spent during our observations, which ranges between $\sim45$~min and $\sim168$~min. Several factors influenced the final ToT for each target, some of which beyond our control (e.g., weather conditions at Paranal or technical issues with the telescope and instrument). To measure the PSF stability, we estimated the flux for each unsaturated PSF frame enclosed in an aperture of $r=3.5$~pix ($\sim1~\lambda/D$) around the image center. The standard deviation of the measured counts normalized to the median count gives a sense of the PSF stability (and thus photometric calibration) of our data. After removing bad frames (see Sect.~\ref{sec:unsat_PSF}), the PSF was very stable in most of the datasets, as shown in the panel {\it (d)} of Fig.~\ref{fig:obs_conditions}.



\section{Data reduction}
\label{sec:data_reduction}
The reduction of our data relies on {\tt PynPoint} \citep{Amara2012, Stolker2019}, an end-to-end pipeline for reduction and analysis of high-contrast imaging data. Two different reduction flows were used for coronagraphic and noncoronagraphic datasets.

\subsection{Preliminary reduction}

\subsubsection{Noncoronagraphic imaging}
\label{sec:non-coro-imaging}
The data reduction for noncoronagraphic data follows that presented in \cite{Stolker2020_pds70}. Briefly, data were corrected for bad-pixels using 4$\sigma$ clipping and substituting the bad pixels with the median of the eight surrounding pixels. The background is removed using the PCA-based algorithm described in \cite{Hunziker2018}, where we benefit from the star hitting a different quadrant of the detector in each cube. Then, images are aligned to each other using a cross-correlation based algorithm and finally centered fitting a 2D Gaussian to the mean of all the images and shifting each image to locate the star at the very center. At this point, the images are cropped to a size that depends on the angular extent of the disk (Table~\ref{tab:target_list}) as detailed in Sect.~\ref{sec:disks}. Finally, we computed the counts in an aperture placed on the star of radius 1 pix and we discarded the frames whose values are more than $1-2~\sigma$ away from the mean depending on the stability of the PSF. This usually resulted in discarding $<10\,\%$ of the frames for each dataset. To reduce the amount of frames and frame-to-frame variations but at the same time keep enough features and diversity in the data, we always averaged over 20 consecutive frames in all datasets.

\subsubsection{Coronagraphic imaging}
The data reduction for coronagraphic imaging follows that presented in \cite{Cugno2019_b}.  The initial ten frames of each cube suffered from a systematic offset that decreases exponentially to a constant level during the sequence \citep[e.g.,][]{Stolker2019} and are therefore discarded to avoid biases during the background subtraction step. The remaining data are first corrected for bad-pixels as above, and subsequently the central star is identified with a 2D Gaussian. In some instances the fit failed, mainly because the AO loop briefly opened. Those frames are easily identified and removed by the frame selection routine later. Detector stripes are removed substituting bad pixels with the mean of the left and right pixels. Then, the background is removed using the median-collapsed offset sky cube taken after every 13 on-target cubes. When possible, the median between the previous and the next sky cubes is used. After background subtraction, the images are centered using the shifts previously registered, when we fit the PSF with 2D Gaussian models. Finally, the images are cropped in size based on the disk extent reported in Table~\ref{tab:target_list} and go through the same frame selection process. Again, the images were binned together every 20 frames.

\subsubsection{Unsaturated PSF}
\label{sec:unsat_PSF}
The unsaturated images were obtained following the same observing and reduction strategy used for noncoronagraphic imaging data, in which the star was placed on three detector quadrants and the two empty quadrants were used to model and subtract the background. After undergoing the same frame selection as in Sect.~\ref{sec:non-coro-imaging} (strictly removing frames further than 1$\sigma$ in this case), the images were median-combined to form a PSF model that could be used to calibrate the data.

\subsection{PSF-subtraction}
The stellar PSF was removed using full-frame principal components analysis \citep[PCA, ][]{Amara2012, Soummer2012} as implemented in {\tt PynPoint} using the median to combine the PSF-subtracted frames. Several hyperparameters could influence the final residuals, such as the number of the removed principal components (PCs) and the central mask applied to cover the very central pixels. To be able to study the innermost regions of protoplanetary disks, we applied a very small mask at the center of the image ($r=0\farcs05$, i.e., $\sim2$ pixels). Furthermore, we always removed between 1 and 40 components and inspected all the images. In this way are able to evaluate more and less aggressive PCA setups, as different reductions are optimized for different regions of the images. For example, the same number of PCs might induce strong self-subtraction at small separations, while at the same time it might not remove enough stellar residuals at larger separations to reveal faint companions in protoplanetary disks. We note that, overall, most of the stellar signal was already removed after 20 components, leaving very clean residuals. 

Bright companions, as is the case for binary systems, might dominate the PCs, leading to a poor PSF subtraction unable to reveal close-in planets. For this reason, an annular mask with a width of $4~\lambda/D$ at the separation of the binary star was applied, in order for the PCA to neglect the stellar companion and to focus on the region close to the primary. 

We visually inspected all the images, searching for faint point sources from young companions. For nine objects, the residuals revealed bright binary system companions that required masking as described above. Once point sources were identified in the residuals, we proceeded with their characterization.

\subsection{Companion characterization}
\label{sec:companion_characterization}
We used two different methods in order to measure the astrometry and the photometry of companion candidates. The fifth column of Table~\ref{tab:companion_candidates_results} indicates which of the two methods was used to infer the properties of the companions.

For point sources in the speckle dominated region of the images ($\rho\lesssim1\farcs0$), we used the MCMC sampling algorithm provided in {\tt PynPoint}, which inserts artificial negative copies of the unsaturated PSF in the images prior to the PCA PSF-subtraction step. Then, the central star was removed, and the residuals at the position of the companion were evaluated in an aperture of size $r\approx~2\lambda/D$ following the method described in \cite{Wertz2017}. The posterior distribution for the three parameters separation, position angle (PA), and contrast was sampled with 300 walkers undergoing chains of 500 steps. For each walker, the first 100 samples are discarded as burn-in phase. Then, the best-fit companion is removed from the images and additional sources with the same contrast are inserted at the same separation as the original one, but with 360 different position angles. Those artificial companions are retrieved so that we could estimate potential biases in our measurement methods. Thus, we correct for the aforementioned bias and we added in quadrature the standard deviation of the 360 retrieved values to the measurement uncertainty. A more detailed description of this approach can be found in \cite{Stolker2020_miracle}.

For companions with $\rho\gtrsim 1\farcs0$ we ran classical ADI (cADI) to reduce self-subtraction and fitted the companion PSF with a 2D Gaussian function. The peak position is used to determine the astrometry of the companion, while the contrast is estimated comparing the amplitude of the fit with the amplitude of the unsaturated PSF after correcting for differences in exposure times. In this case we conservatively considered fixed uncertainties of 9 mas for both RA and DEC coordinates, as this was the total uncertainty on the separation that we measured for HD101412\,C (see Table~\ref{tab:companion_candidates_results} and Sect.~\ref{sec:HD101412}) and it is unlikely that the Gaussian fitting method carries a larger uncertainties than those measured in the speckle dominated region at $0\farcs17$. The uncertainties on the contrast are calculated assuming an error equal to the variability of the PSF (Table~\ref{tab:observations}).

Finally, for both methods separations and position angles were corrected using the plate scale and the true north corrections estimated in \cite{Launhardt2020}. We report the final values in Table~\ref{tab:companion_candidates_results}.

\begin{table*}[t!]
\centering
\caption{Full list of stellar and substellar companions detected in the NaCo-ISPY PPD sample.}
\def\arraystretch{1.25}
\small
\begin{tabular}{lllllllllll}
Target      & Sep ($^{\prime\prime}$)& Proj. sep. (au) & PA ($^\circ$) & $\Delta L'$   & Method        & Other reference\\ \hline
HD 35187 B      & $1.366\pm0.013$   & $222.4\pm2.8$    & $193.8\pm0.2$ & $2.3\pm0.1$   & Gaussian fit  & \cite{Dunkin1998}    \\
HD 37411 B      & $0.456\pm0.013$   & $161.0\pm5.3$    & $356.7\pm0.2$ & $1.7\pm0.1$   & Gaussian fit  & \cite{Thomas2007}  \\
HD 37411 C      & $0.579\pm0.013$   & $204.4\pm5.7$    & $349.8\pm0.2$ & $2.0\pm0.1$   & Gaussian fit  & \cite{Thomas2007}  \\
HD 38120 B      & $1.265\pm0.013$   & $490.1\pm8.1$    & $128.6\pm0.2$ & $0.4\pm0.1$   & Gaussian fit  & $-$ \\
V* NX Pup B     & $0.182\pm0.017$   & $75.2\pm16.6$    & $86.1 \pm1.7$ & $6.3\pm0.4$   & MCMC          & \cite{Schoeller1996} \\  
HD72106 B       & $0.761\pm0.013$   & $293.3\pm10.0$   & $29.4 \pm0.2$ & $-0.1\pm0.1$  & Gaussian fit  & \cite{Vieira2003} \\
HD 100453\,B    & $1.076\pm0.013$   & $112.5\pm1.4$    & $129.6\pm0.2$ & $6.1\pm0.2$   & Gaussian fit  & \cite{Collins2009}  \\
HD 101412\,B    & $0.533\pm0.005$   & $221.2\pm2.0$    & $147.9\pm0.3$ & $8.0\pm0.1$   & MCMC          & Ruh et al. (in prep.)  \\
HD 101412\,C    & $0.170\pm0.009$   & $70.7\pm3.7$     & $181.0\pm1.3$ & $7.4\pm0.2$   & MCMC          & Ruh et al. (in prep.)  \\
HD 104237\,B    & $1.373\pm0.013$   & $147.5\pm0.7$    & $254.0\pm0.2$ & $6.2\pm0.1$   & Gaussian fit  & \citep{Grady2004}      \\
PDS70\,b        & $0.207\pm0.015$   & $23.4\pm1.6$     & $149.8\pm1.7$ & $6.8\pm0.2$  & $-$           & \cite{Stolker2020_pds70}\\
PDS70\,c        & $0.254\pm0.010$   & $28.8\pm1.1$     & $283.3\pm2.0$ & $6.6 \pm 0.2$ & $-$           & \cite{Haffert2019}\\
HD144432 BC     & $1.490\pm0.013$   & $232.4\pm2.2$    & $5.9  \pm0.2$ & $1.6\pm0.1$   & Gaussian fit  & \cite{Muller2011} \\
V* KK\,Oph\,B   & $1.602\pm0.013$   & $269.6\pm6.4$    & $244.1\pm0.2$ & $2.5\pm0.1$   & Gaussian fit  & \cite{Leinert1997}   \\
V* R\,CrA\,B    & $0.187\pm0.006$   & $23.4\pm1.6$     & $132  \pm0.7$ & $6.7\pm0.2$   &  $-$          & \cite{Cugno2019_b}      \\\hline
\end{tabular}
\tablefoot{If a method is reported, separation, position angle, and contrast were calculated as detailed in Sect.~\ref{sec:companion_characterization}. If a method is not reported, the values of the astrometric and photometric parameters were taken from the reference reported in the last column, as the same dataset was already presented in those papers.}
\label{tab:companion_candidates_results}
\end{table*}

\subsection{Contrast curves}
\label{sec:contrast_curves}
We calculated contrast curves for all data sets in the survey using {\tt applefy}, 
which follows the routine presented in \cite[submitted]{Bonse2023}. Our analysis starts with a preparatory examination of residual noise statistics. For this purpose, we compute Q-Q plots to compare the pixel noise in the residuals with Gaussian noise. Exemplary results for HD31648 and HD36112 are given in App.~ \ref{app:qqplots}. The residual noise in areas which do not contain extended scattered light signals from a protoplanetary disk is mostly consistent with Gaussian noise. Although Q-Q plots cannot prove that the actual noise originates from a Gaussian distribution, we assume that the noise is sufficiently normal to perform a t-test \citep{Mawet2014}. On the contrary, in areas of our images which contain extended scattered light from a protoplanetary disk the statistic is dominated by the disk signal. In those regions, the noise is not normal (see right panel of Fig.~\ref{fig:QQplots}), and the noise can neither be considered independent (due to the extended nature of the disk signal) nor identically distributed (given that some areas are related to the disk signal, some to the dark regions originating from self-subtraction, and some are estimates of the true speckle and detector noise). Hence, none of the assumption necessary to compute detection limits based on the t-test \citep{Mawet2014} is applicable. As a consequence, we exclude these regions from our analysis (usually the first few $\lambda/D$). This problem is limited to the innermost region of 17 targets (see Sect.~\ref{sec:disk_det}), while the outer disk is never detected in scattered light in $L'$. For these 17 sources we started calculating contrast curves from the first fixed separation not including disk signals (see below) and we ignored regions at smaller separations. 

The calculation of the contrast curves relies on three main components: the detection threshold, the signal of the planet and the strength of the noise. We fix the detection threshold to a false-positive-fraction (FPF) of $2.87 \times 10^{−7}$ for all separations, which is equivalent to $5 \sigma$ for large separations. We inject fake planets at different separations from the star (steps of $1 \lambda /D$ at separations lower than 12 $\lambda /D$, steps of $3 \lambda /D$ at larger separations as the contrast is expected to stabilize) using the unsaturated PSF. In this way we can calculate the attenuation of the planetary signal caused by over- and self-subtraction during the data post-processing. In order to account for azimuthal variations, 6 planets (one planet at a time) are inserted for every separation. The result is the throughput of the data post-processing as a function of separation averaged over 6 azimuthal positions. In contrast to the metric presented in \cite{Mawet2014}, we do not use apertures but pixel values spaced by 1 FWHM \citep[submitted]{Bonse2023}. In this way, the noise is approximately uncorrelated, which is a prerequisite for use of the t-test. For every separation we extract the noise for 360 different placements and report the median over all results.

\section{Results on individual objects}
\label{sec:individual_objects}

Residuals on each individual target are shown in Figs.~\ref{fig:residuals_1} through \ref{fig:residuals_3}. There, companions and concrete candidates are highlighted with circles, while nondetections of known companions at locations predicted by previous studies are marked with dashed circles. In Sect.~\ref{sec:new_companions} we focus on newly detected companions, in Sect.~\ref{sec:undetected_companions} we introduce interesting upper limits on individual undetected companions whose presence has been inferred with both direct and indirect methods, and in Sect.~\ref{sec:disk_det} we present the detection of disk signals. In App.~\ref{sec:known_companion} we describe the detection of known stellar and substellar companions, while in App.~\ref{sec:bkg_obj} we report the detection of background objects. A list of all the detected companions, together with their properties can be found in Table~\ref{tab:companion_candidates_results}.

\subsection{Newly detected companions around individual objects}
\label{sec:new_companions}

\subsubsection{HD38120}
\begin{figure}
    \includegraphics[width=0.98\hsize]{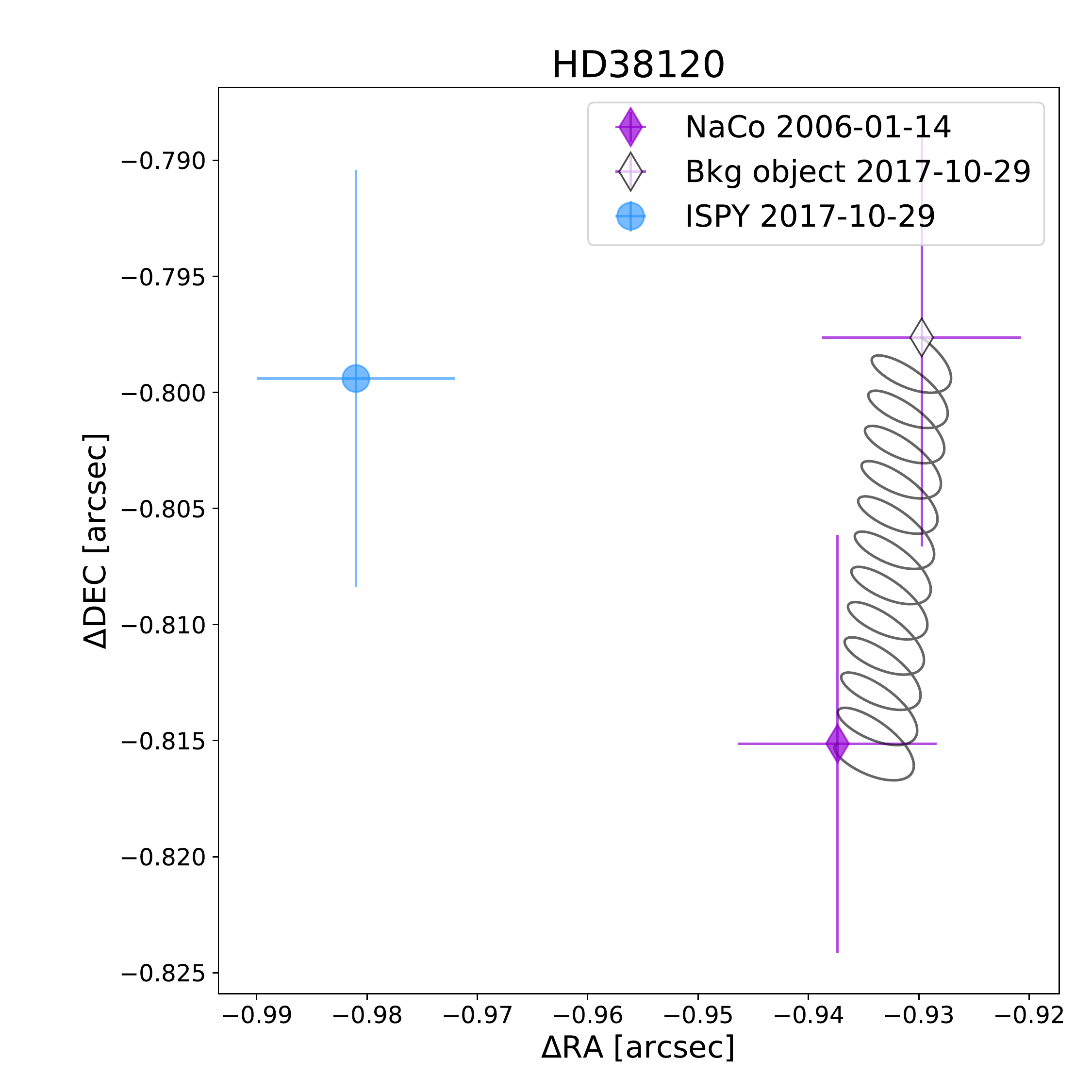}
    \caption{Proper motion analysis of the stellar companion candidate orbiting HD38120. The companion motion is inconsistent with a background object, and consistent with being bound. The dark diamond and circle are the 2006 and 2017 positions respectively, the solid line is the expected motion of a background object and the faded diamond is the expected position of a background object in 2017.}
    \label{fig:proper_motion_analyisis}
\end{figure}

The image of HD38120 shows residuals from a companion candidate at the edge of the field of view (Fig.~\ref{fig:residuals_1}). For this reason, we enlarged it by $0\farcs5$ in radius, to include the signal coming from the potential companion. The companion candidate is detected at a separation of $1\farcs265$ with a contrast of 0.4~mag. Having similar brightness in the $L'$ images, it is unlikely that the companion candidate is a background object. Nonetheless, we reduced archival NaCo data in the $K_s$ band (Prog. ID: 076.C-0679(B), PI: Bouwman) taken in 2006 in order to verify that the two objects are comoving. The proper motion analysis is reported in Fig.~\ref{fig:proper_motion_analyisis}, showing that the candidate motion is inconsistent with a background object.

We further test the binary scenario by checking whether the candidate motion is consistent with a bound object, using the method of \cite{Pearce2015}. Their parameter $B$ (Eq. 1 in that paper) combines sky-plane separation, relative velocity and mass to assess whether two objects can be bound; if ${B<1}$ then bound companionship is possible (although not certain, because the line-of-sight coordinates are unknown), whilst if ${B\geq1}$ then companionship is ruled out because the relative velocity would be too high (regardless of line-of-sight coordinates). We calculate $B$ for HD38120 using the 2006 and 2017 data (separated by ${11.8 \; \rm yr}$), for which the separations are ${1\farcs242 \pm 0\farcs009}$ and ${1\farcs265 \pm 0\farcs013}$ and the position angles ${130.4 \pm 0.4^\circ}$ and ${128.6 \pm 0.4^\circ}$ respectively (assuming uncertainties of one-third of a pixel, see Sect.~\ref{sec:companion_characterization}). We use a primary mass of ${2.6 \pm 0.1 \; M_\odot}$, and assume that the secondary has the same mass with a ${100 \; \rm per cent}$ uncertainty (since the contrast in L$'$ is only \mbox{0.4 mag}). These yield ${B = 2.6 \pm 2.0}$; it is therefore possible that ${B<1}$ (within the uncertainties), and so it is dynamically possible for the pair to be bound. Reducing the uncertainty on the secondary mass would not change this conclusion.

\subsubsection{HD101412}
\label{sec:HD101412}
Two new companions have been discovered around HD101412 in our $L'$ images within $0\farcs6$ from the primary. The study of the proper motion analysis, together with the companions classification performed with multiwavelength follow-up observations will be presented in a separate manuscript (Ruh et al., in prep.), while here we only report astrometric and photometric properties measured in the $L'$ band. We note that \cite{Rich2022} detected the same objects in GPI $H$ band data, and they assessed with 3$\sigma$ confidence that the point sources are not background objects, in agreement with Ruh et al. (in prep.).

\begin{figure*}[h!]
    \centering
    \includegraphics[width=0.9\hsize]{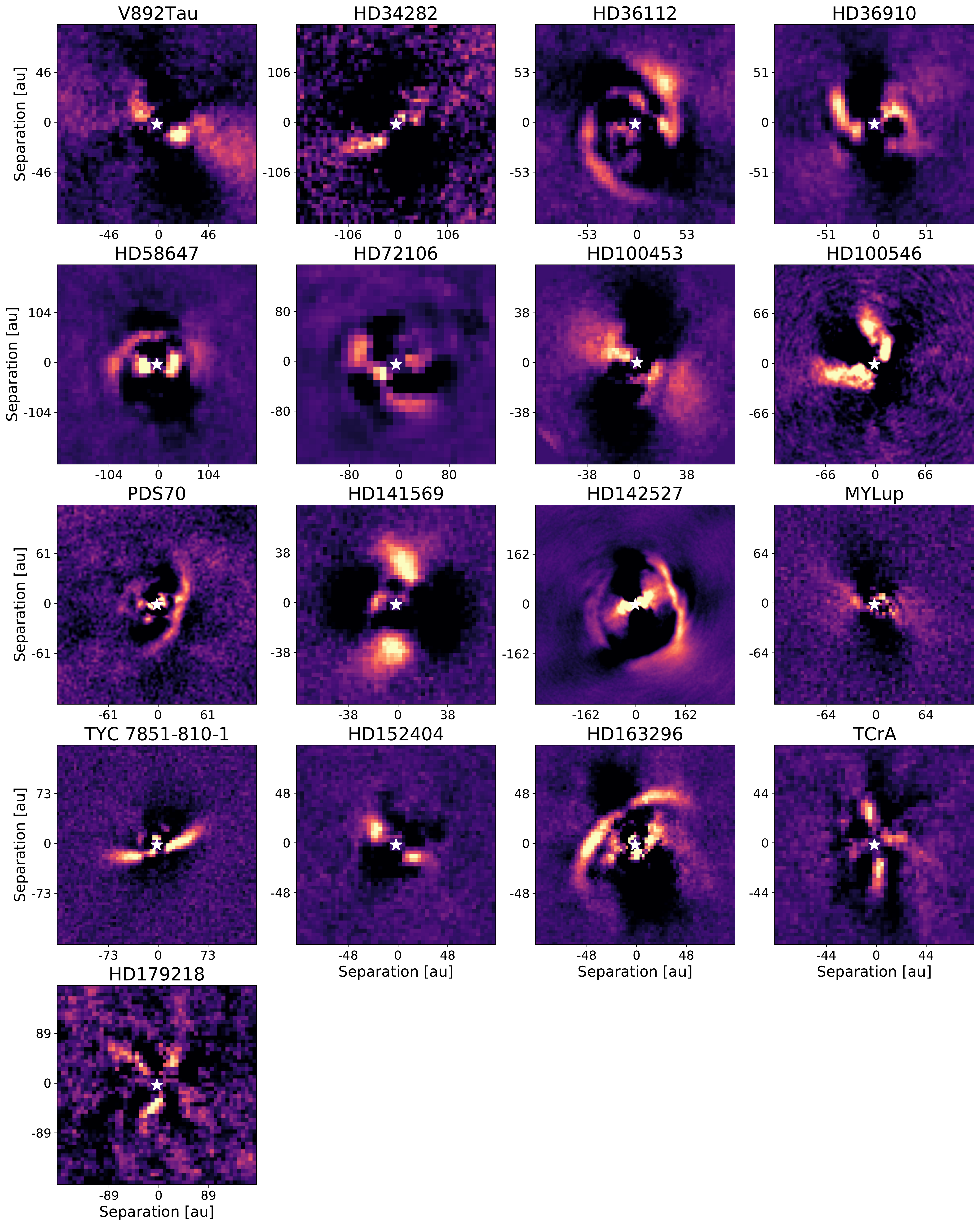}
    \caption{NaCo-ISPY $L'$ gallery of detected protoplanetary disks. Images have been cropped to highlight the disk morphology and the PSF-subtraction parameters were adapted to show the brightest possible disk residuals.}
    \label{fig:disks}
\end{figure*}

\begin{figure*}[h!]
    \centering
    \includegraphics[width=0.95\hsize]{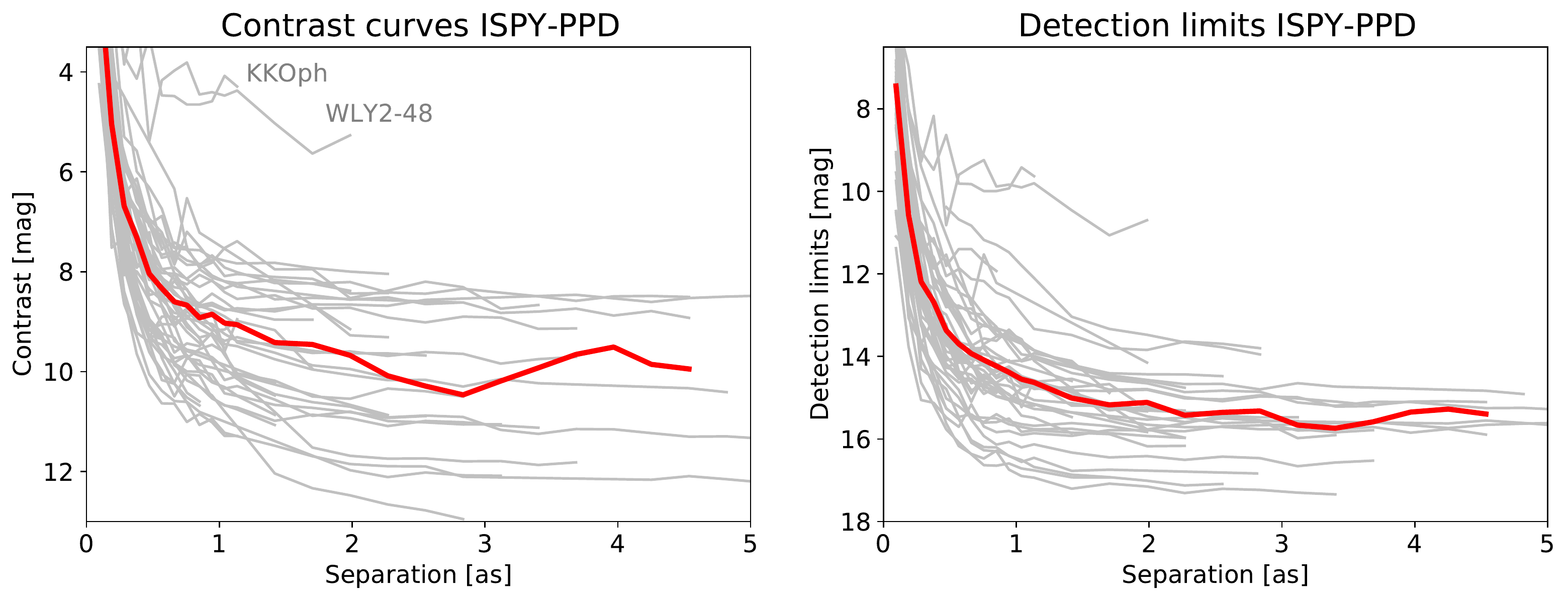}
    \caption{Contrast limits ({\it left}) and apparent magnitude detection limits ({\it right}) estimated for the ISPY PPD sample as a function of separation. Gray lines represent limits for the individual targets, while the red thick line reports the median limit calculated at each separation. All the curves are obtained for a FPF$=2.87\times10^{-7}$.}
    \label{fig:Contrast_curves}
\end{figure*}

\subsection{Interesting nondetections and dubious candidates}
\label{sec:undetected_companions}

\subsubsection{HD97048}
\cite{Pinte2019} identified a kink in the isovelocity curve of HD97048 at a separation corresponding to one of the dust gaps from the disk ($\rho=0\farcs45\pm0\farcs1$, PA~$=-55\pm10^\circ$, $\Mp=2-3~\MJ$). The NaCo images do not reveal a signal corresponding to that position, at which we reached a contrast of $\sim7.9$~mag. Using the information provided in Table.~\ref{tab:target_list}, we estimate a mass upper limit of $\sim52~\MJ$ using the Ames-Dusty evolutionary models \citep{Chabrier2000}. Much deeper observations will be necessary to unveil the companion in the dust gap.

\subsubsection{HD100546}
\cite{Quanz2013} claimed the detection of a protoplanet $\sim47 \pm 4$~au away from the Herbig Ae/Be star HD100546 using $L'$ data from VLT/NaCo. They confirmed the detection with a second independent $L'$ dataset \citep{Quanz2015}, detected it in the $M'$ band as well, and used $K_s$-band data to put an upper limit on the companion flux at shorter wavelengths. Fitting the few available datapoints, \cite{Quanz2015} concluded that they detected emission from the hot circumplanetary environment rather than from b itself. Furthermore, \cite{Currie2014} and \cite{Currie2015} identified the planet b in GPI $H$-band data, confirming the very red IR colors, as expected for an embedded object. More recent works cast doubts on the existence of the protoplanet, suggesting that the detections are the result of scattered light from the disk after aggressive post-processing \citep{Garufi2016, Follette2017, Rameau2017}. Finally, \cite{Mendigutia2017} and \cite{Cugno2019_a} searched for H$\alpha$ signals emitted from the accretion shock surface without finding any.

The ISPY data revealed a potential point source at the position of HD100546\,b, best visible with relatively aggressive reductions obtained with a high number of principal components.  
However, the feature seems to have an elongated shape and it sits on a bright disk arm, as noted by \cite{Quanz2013} and \cite{Quanz2015}. Given the rather debated nature of HD100546\,b,  this dataset will be studied in a separate paper focused uniquely on this object, and we do not consider it to be a confirmed companion in this study. 

\subsubsection{HD142527}
HD142527 hosts a disk with a very large optically thin cavity. An accreting stellar companion is located within the cavity \citep{Biller2012, Close2014} at $\sim0\farcs063$ \citep{Cugno2019_a, Balmer2022} on an orbit misaligned with the outer disk \citep{Lacour2016, Balmer2022}. Being so close to its host, HD142527\,B falls at a separation smaller than the angular resolution of our NaCo observations ($\lambda/D\approx0\farcs095$). Hence, we did not detect it.

\subsubsection{HD163296}
\cite{Pinte2018} inspected channel maps of the disk around HD163296, identifying a kink in the velocity field of the disk gas, presumably caused by the presence of a forming companion with mass $\sim2~\MJ$. The planet was later independently confirmed by \cite{Teague2021} and \cite{Izquierdo2021} at a separation of $\sim2\farcs0$ with PA$\sim0^\circ$. Our ISPY observation did not directly show the embedded companion, and at the expected companion separation we reached a contrast of 11.8~mag, corresponding to a limit on the brightness of the planet of 15.3 mag. Assuming that the emission is due to photospheric emission only (no accretion) and that the disk material does not influence at all the emitted flux, we can compare this value with the expected value from the hot-start Ames-Dusty evolutionary models \citep[as used by \citealt{Asensio-Torres2021}]{Chabrier2000} for a planet coeval with the parent star (7.1 Myr, see Table~\ref{tab:target_list}). In this framework of assumptions, the detection limits exclude planet masses larger than $\sim4.2~\MJ$, consistent with our nondetection of HD163296\,b. We also note that our contrast curve is roughly consistent with that obtained by \cite{Guidi2018} with Keck/NIRC2 after correcting for the different statistical significance. Deeper observations, potentially with JWST, are necessary to confirm the indirect detection by \cite{Pinte2018}.

\subsection{Disks}
\label{sec:disk_det}

Even if our observational strategy and reduction pipeline were not optimized for the detection of protoplanetary disk signals, 17 protoplanetary disks could be detected in the final residuals out of the 45 targets we observed with the VLT/NaCo instrument. For all these sources, the disk detections are reported in Fig.~\ref{fig:disks}. Most of the disks from Fig.~\ref{fig:disks} were already known and images were taken in the past either with high-contrast imagers or with the ALMA observatory. 
Some of these disks were also already imaged in the $L'$ band. These are HD34282 \citep{Godoy2021, Quiroz2022}, HD36112 \citep{Reggiani2018, Wagner2019}, HD36910 \citep{Uyama2020}, HD100453 \citep{Wagner2018}, HD100546 \citep{Quanz2013, Quanz2015}, PDS70 \citep{Keppler2018, Wang2020, Stolker2020_pds70}, HD141569 \citep{Mawet2017}, HD142527 \citep{Rameau2012}, and HD163296 \citep{Guidi2018}. 

For other targets the first $L'$ disk images are reported in this work (some were already introduced in \citealt{Launhardt2020}, as part of an introduction to the ISPY survey). These are HD58647 and V892 Tau (\citealt{Stapper2022} presented ALMA images for these sources), MY~Lup (previously imaged by \citealt{Avenhaus2018} with VLT/SPHERE in polarimetric mode), TYC 7851-810-1 (imaged with ALMA by \citealt{Ansdell2018}), HD152404 (\citealt{Janson2016} showed SPHERE/IRDIS images of the disk) and HD179218 (see also \citealt{Kluska2018} and their VLT/SPHERE observations). Finally, to our knowledge this is the first direct detection of the disks surrounding HD72106 and T~CrA.  

A coherent and exhaustive analysis of the disk images involves the study of the disk at other wavelengths, coupled with a radiative transfer code and assumption on the dust and gas distribution and properties. This is beyond the scope of this work, and we leave the interpretation of the disk signals to future work. 
However, in the next paragraph, we use the ellipse fitting tool from \cite{Hammel2020} to derive at least some of the basic parameters of the newly discovered disks, especially the inclination that we subsequently use in Sect.~\ref{sec:completeness}. 

First, we produce radial profiles in every $3^\circ$ azimuth section and find the peak positions of the ring-like emission. To compensate for the coarse sampling of the disk near the major axis we sample every $1^\circ$ near that region. The peak pixel coordinates are provided as input to the ellipse fitting routine. For each target, we perform the fitting on multiple images obtained by subtracting different numbers of components. The final values result from the average and the standard deviation of these separate fittings. The inclination is estimated from the aspect ratio assuming that the disk is a circle if seen face-on.

Given the geometrical similarities between the disks around HD72106 and T\,CrA and the disk around TYC~7851-810-1, we used the latter to verify our procedure. The fit results $i=72\pm 3^\circ$, PA = $105\pm 2^\circ$ are consistent with literature values $i=74^\circ$ and PA =$107^\circ$ from \cite{Ansdell2018}.
We derive for HD72106 a disk inclination $i=51\pm 4^\circ$ and for T\,CrA $i=77\pm 2^\circ$. In addition, we obtain the position angle of the major axis PA = $47\pm 2^\circ$, and ring radius $R=70\pm 2$ au for HD72106, PA = $5\pm 2^\circ$, and $R=35\pm 2$ au for T\,CrA. Inclinations are reported in Table~\ref{tab:target_list}.

\begin{figure}
    \centering
    \includegraphics[width=\hsize]{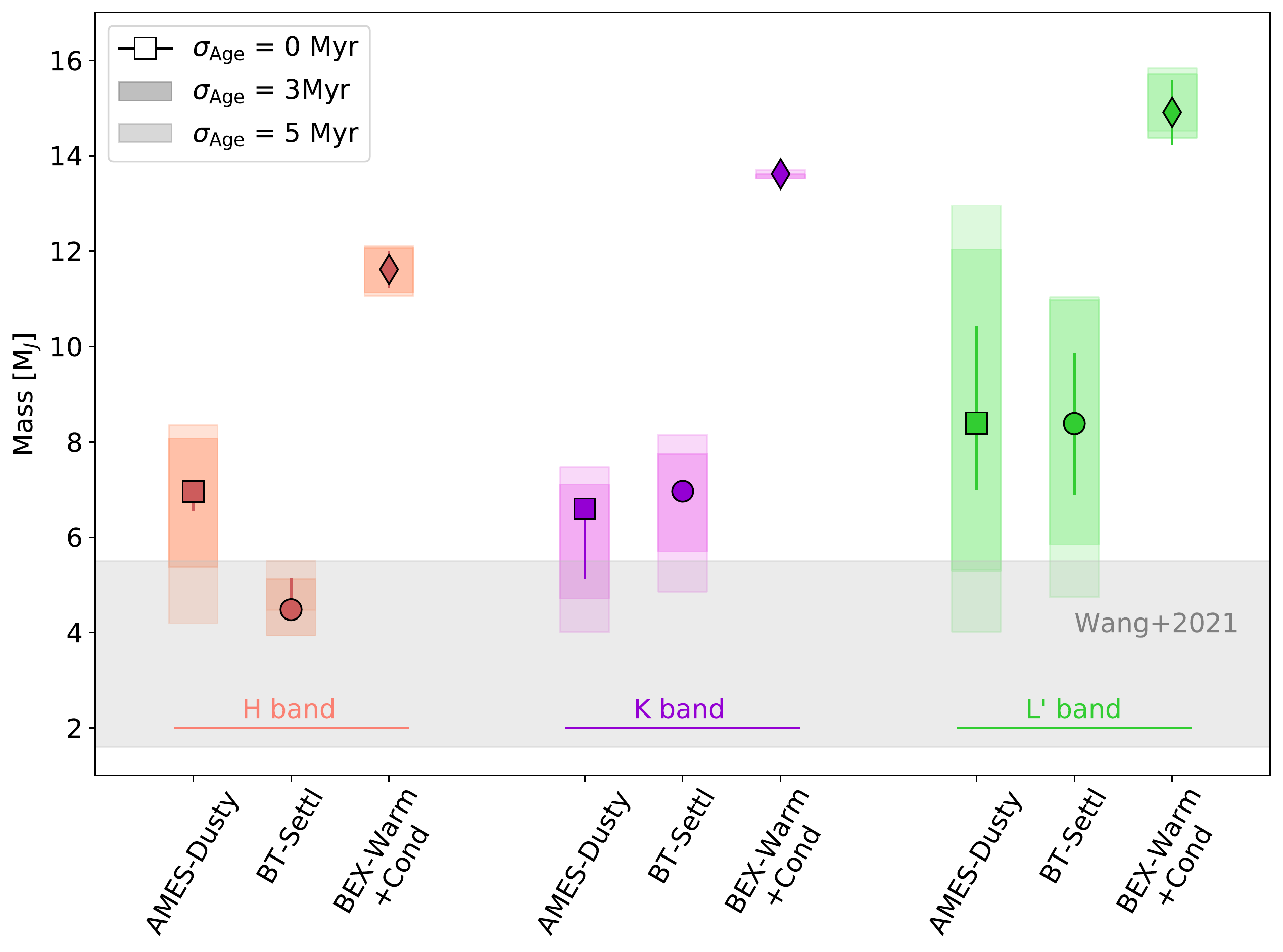}
    \caption{Planetary mass estimate as a function of the assumed evolutionary track, atmospheric model and age uncertainties for the PDS70b protoplanet. Colors represent the different bands considered here and are reported at the bottom of each band. Markers and their errorbars represent the mass estimates when the age is assumed to be known and exact ($\tau=8$~Myr, $\sigma_\mathrm{Age}=0$~Myr) when using AMES-Dusty (squares), BT-Settl (circles) and BEX-Warm+Cond (diamond). Shaded areas represent the same measurement considering different uncertainties ($\sigma_\mathrm{Age}=3$~Myr for the more intensely colored regions, $\sigma_\mathrm{Age}=5$~Myr for the more transparent regions). The gray region represents the 68\% confidence level range of the value for dynamical mass of PDS70b calculated in \cite{Wang2021} thanks to the astrometric precision of the VLTI/GRAVITY instrument. }
    \label{fig:mass_PDS70b}
\end{figure}

\section{Results on the overall sample}
\label{sec:results_all}

In this Section we aim to interpret the results of our survey as a whole, drawing conclusions that can statistically constrain the population of forming planets. In Sect.~\ref{sec:contrast_sample} we look at the detection limits of our targets, in Sect.~\ref{sec:mass_problem} we lay out several issues related to the transformation of detection limits into mass upper limits and we propose a solution in Sect.~\ref{sec:teff_limit}. Finally, in Sect~\ref{sec:completeness} we compute completeness maps for the ISPY PPD survey. 

\subsection{Contrast and detection limits for the ISPY PPD sample}
\label{sec:contrast_sample}
Figure~\ref{fig:Contrast_curves} shows in gray the contrast curves obtained by applying the procedures described in Sect.~\ref{sec:contrast_curves} on each of the datasets presented in this paper. In addition, thick red lines show the median contrast obtained at each separation. The contrast performance of the WLY2$-$48 and KK\,Oph datasets are much worse than for the rest of the targets due to problems with the AO loop stability occurring during the observations (WLY2$-$48) and the presence of an equal brightness companion in the image that dominates the residuals (KK\,Oph). 
As discussed in Sect.~\ref{sec:contrast_curves}, each contrast curve might have a different starting separation depending on the presence of a bright disk in scattered light preventing a reliable quantification of the $\mathrm{FPF}=2.87\times10^{-7}$ contrast at small separation, and a different radial extent depending on the radius of the protoplanetary disk surrounding the target. Thus, the number of contrast curves contributing to the median estimate at each separation may vary. 

Overall, the left panel of Fig.~\ref{fig:Contrast_curves} shows that we reached median contrasts of 6.1, 8.1, 9.0 and 10.2 mag at separations $\rho = 0\farcs25, 0\farcs5, 1\farcs0 $ and $2\farcs0$, with a general scatter of $\sim1.5-2$ mag on both sides. This is roughly in line with the values found by \cite{Launhardt2020} for the preliminary analysis of the entire NaCo-ISPY survey, despite the fundamentally different methods employed for the estimate of the contrast limits. 

Detection limit curves were obtained adding to each curve the apparent $L'$ magnitude of the star. After this operation, the spread of the curves, especially in the background limited regime, is much smaller. We reached a median detection limit of 11.6, 13.5, 14.5 and 15.4 mag at separations $\rho = 0\farcs25, 0\farcs5, 1\farcs0 $ and $2\farcs0$.

\begin{figure*}[t!]
    \includegraphics[width=\hsize]{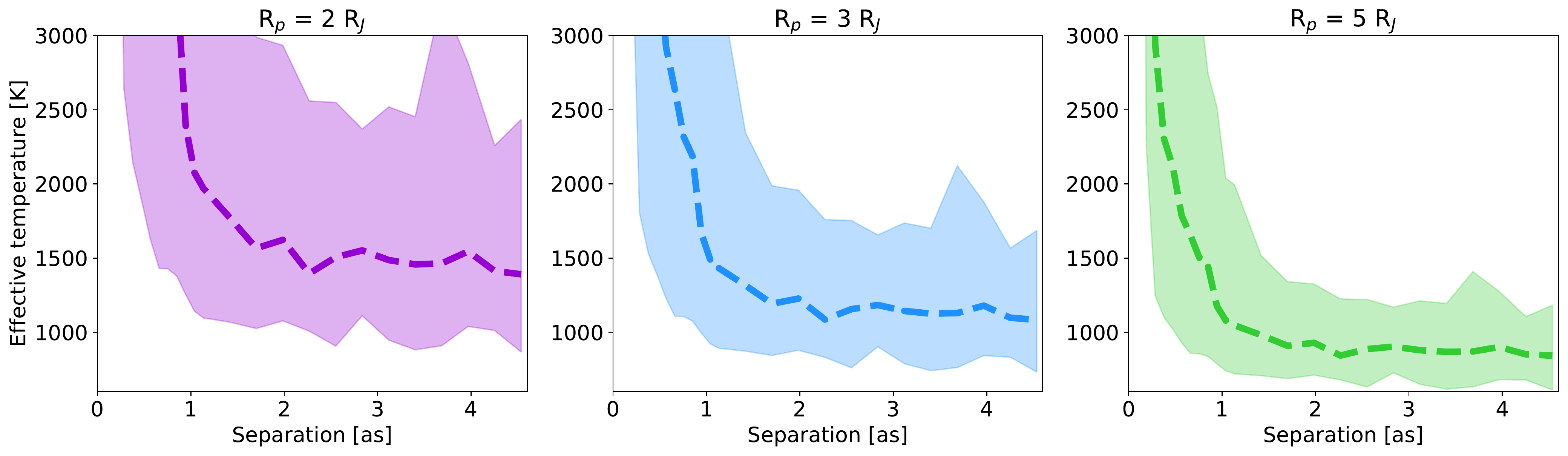}
    \caption{Effective temperature limit of the whole NaCo-ISPY PPD sample assuming $\Rp=2\RJ$ (left), $\Rp=3\RJ$ (middle) and $\Rp=5\RJ$ (right). The dashed line represent the median temperature for the targets having limits extending up to that separations, while the shaded area represents the 16-84\% range.}
    \label{fig:Teff_curves}
\end{figure*}

\subsection{The problem of the mass-luminosity conversion}
\label{sec:mass_problem}
Most of the high-contrast imaging surveys run in the past used age estimates together with evolutionary and atmospheric models to transform flux detection limits into mass upper limits, therefore being able to constrain the planet population potentially detectable by the observations \citep[e.g.,][to name a few]{Stone2018, Nielsen2019,Vigan2021}. For forming planets, such an approach strongly relies on several assumptions: (i) atmospheric model, (ii) evolutionary model, (iii) age estimate and age uncertainty, (iv) presence of accretion processes, and (v) extinction along the line of sight. Before proposing an alternative approach in Sect.~\ref{sec:teff_limit}, we discuss each of those points with the help of Fig.~\ref{fig:mass_PDS70b}, which shows the magnitude-to-mass conversion for the measured photometries in the $H$, $K$ and $L'$ bands of the protoplanet PDS70\,b taken from \cite{Stolker2020_pds70} assuming an age of $\tau=8$~Myr \citep{Wang2021}, with uncertainty $\sigma_\mathrm{Age} = 0$~Myr unless stated otherwise.

The choice of atmospheric model used to describe the planet emission might influence the interpretation of the detection limits. For example, different cloud treatments or varying the opacity sources could change the flux in every band. In Fig.~\ref{fig:mass_PDS70b} the mass estimated from the $H$ band measurement for the AMES-Dusty \citep{Chabrier2000} and BT-Settl \citep{Baraffe2015} model vary by a factor 1.6 (those models assume similar initial entropy following ``hot start'' scenario). Conversely, the masses estimated from $K$ and $L'$ photometry seem to be consistent with each other for the two models.

The choice of evolutionary model strongly impacts the emission of substellar objects, especially at young ages (see for example \citealt{Spiegel2012}). In such cases, assuming a hot-, a warm- or cold-start model can strongly bias the final results. 
Figure~\ref{fig:mass_PDS70b} shows the mass estimate for hot-start isochrones (AMES-Dusty, BT-Settl) and warm-start models (BEX-warm, \citealt{Marleau2019}). The warm start models need more massive objects to match the brightness measured for PDS70\,b, with a factor $\sim1.7-2.6$ difference between the mass values estimated by hot and warm evolutionary tracks.

Age estimates of young stars strongly depend on the method used for the derivation, and different methods very often deliver very different results.
Furthermore, depending on the planet formation model, there might be a delay between the time planets and stars start their lives. In Fig.~\ref{fig:mass_PDS70b} we report the mass uncertainities for $\sigma_\mathrm{Age} = 3$~Myr and $\sigma_\mathrm{Age} = 5$~Myr as shaded regions (see legend). Especially in the $L'$ band, the ratio between maximum and minimum mass range is 2.3 (3.2) for $\sigma_\mathrm{Age}=3~(5)$~Myr.

Accretion processes from the protoplanet environment onto the circumplanetary disk (CPD) and the planet surface may substantially increase the observed luminosity \citep[e.g.,][]{Szulagyi2014, Zhu2015, Szulagyi2017}. As a consequence the direct conversion of the measured flux to mass could lead to a biased mass estimate. Finally, extinction from circumstellar (particularly for non face-on disks) and circumplanetary material could influence the emission able to escape the protoplanetary disk, again impacting our ability to convert photometric flux measurements and detection limits into masses \citep{Szulagyi2018, Sanchis2020}.

From these arguments we can understand that the problem of converting flux measurements into masses is extremely challenging for young forming planets and quickly becomes degenerate. Thus, it is almost impossible to actually constrain the population of forming gas giant planets using the mass as a key population parameter. 
In particular in the $L'$ band, the uncertainties related to all these factors suggest that any assessment of the presence of planets and their mass will depend mostly on the underlying assumptions rather than the detection limits estimated from the data.

As an additional consideration, we overplotted in gray the PDS70~b mass estimate obtained when requiring the PDS70 system to be dynamically stable. Most photometric mass estimates seem to disagree with the measured dynamical mass most likely due to one or a combination of the assumptions above. More data are required to validate these preliminary findings and confirm the dynamical mass of PDS70\,b, but there is the concrete possibility that at very young ages the standard magnitude to mass conversion is not an appropriate tool to constrain the architecture of infant planetary systems. Similar tensions between dynamical mass measurements and mass estimates based on isochronal fitting were also highlighted in the past. For example, \cite{Dupuy2009, Dupuy2014} and \cite{Kuzuhara2022} found a relevant difference between the two values for several brown dwarf binaries and companions. In the next Section we try to overcome this problem using existing information on the spectral emission of the forming planets PDS70~b and c.

\begin{figure*}[t!]
    \includegraphics[width=\hsize]{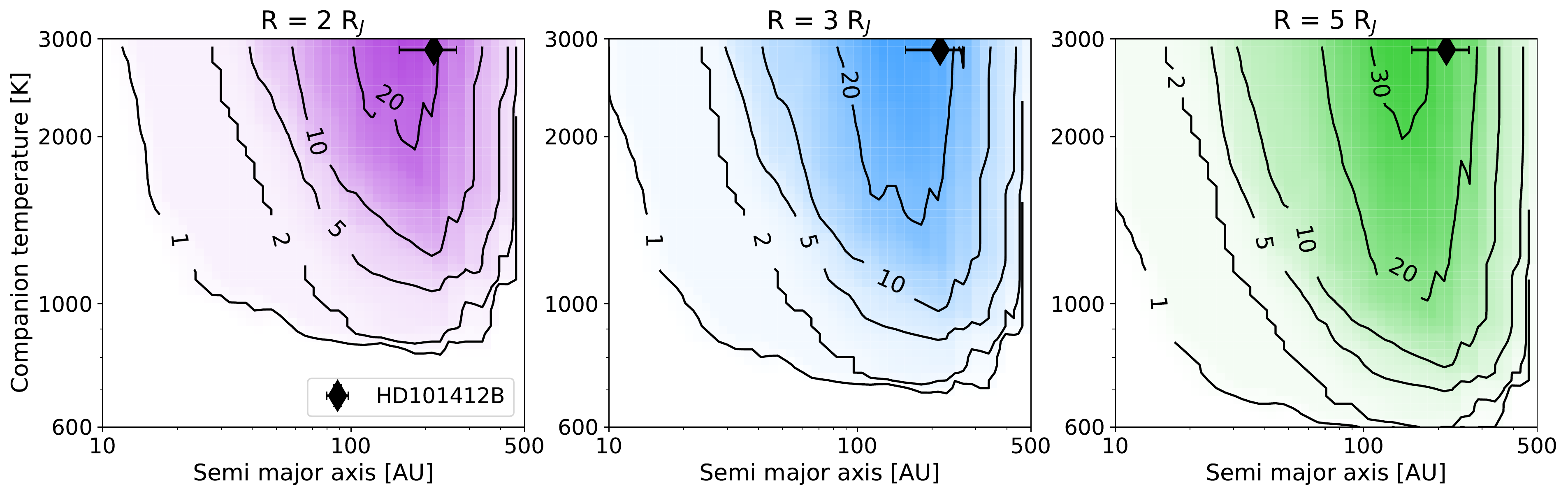}
    \caption{Depth of search for the 45 ISPY-PPD targets included in the analysis, reporting the number of stars to which the survey is complete for young forming planets in the $\Teff-a$ parameter space. The three plots represent maps when assuming $\Rp=2,3,5~\RJ$ ({\it left, mid} and {\it right} panels, respectively). Overplotted as full black marker the companion HD101412 B. }
    \label{fig:prob_map}
\end{figure*}

\subsection{Temperatures as a model- and age-independent parameter to constrain the population of forming planets}
\label{sec:teff_limit}

Pursuing a different approach, we remain as close as possible to the data, obtaining results that are independent from a multitude of arbitrary assumptions. Following recent work on PDS70\,b and c, where the SEDs of the planets was found to be well described by a blackbody function \citep{Wang2020, Stolker2020_pds70, Wang2021} and given the lack of detectable molecular features \citep{Cugno2021}, we convert the detection limits into effective temperatures characterizing black body emission. We considered planet sizes of $\Rp = 2,3,5~\RJ$ and we estimated the effective temperature that generates a black body emission bright enough to be detected by our NaCo observations in the $L'$ filter (the $L'$ flux was estimated using the NaCo $L'$ filter transmission profile and the {\it species} toolkit, \citealt{Stolker2020_miracle}). 

The three planet radii considered here were chosen based on the following ideas: \cite{Ginzburg2019} demonstrated that $\Rp\simeq2\RJ$ during the last few Myr of planet formation in case of low opacity (dust-free) atmospheres for a multi-$\MJ$ planet, while larger radii ($\Rp\approx5\RJ$) can be invoked for young planets whose atmospheres contain considerable amounts of dust. Furthermore, \cite{Stolker2020_pds70} estimated a photometric radius of $\Rp=3\RJ$ for PDS70\,b from its SED assuming blackbody emission. 

We note that we did not consider smaller planets, for example with $\Rp=1~\RJ$, as this would most likely not be a realistic case for a young multi-$\MJ$ planet, as shown by theoretical modeling \citep[e.g.,][]{Spiegel2012, Marleau2014, Mordasini2012} and observations of young directly imaged planets so far \citep{Stolker2020_pds70, Wang2021, Doelman2022, Currie2022_ABAur}. Indeed, during this preliminary phases of their lives, planets are still contracting while emitting a lot of radiation \citep[e.g.,][]{Burrows2001}, and therefore they still appear inflated.

In Fig.~\ref{fig:Teff_curves} we show the median and the 16-84 quantiles of the effective temperature limits as a function of the separation from the star. To estimate those limits we considered at each separation only the targets whose images were large enough (see criterion in Sect.~\ref{sec:targets}) and whose disk scattered light emission contaminated the inner part of the images (see Sect.~\ref{sec:contrast_curves}). As expected, larger planets provide colder limits, while smaller planets could only be detected when hotter. At separations larger than $1\farcs0$, the median temperature limit obtained by our survey is $\Teff=1600, 1200, 900$~K for $\Rp=2,3,5~\RJ$.

\subsection{Survey completeness}
\label{sec:completeness}

To assess the completeness of our survey, we used Monte-Carlo (MC) simulations \citep{Kasper2007, Nielsen2008} evaluating the detection probability over a grid uniform in log space in effective temperature (range $600-3000$~K, 50 steps) and semi-major axis (range $10-500$~au, 50 steps).
To each ($\Teff$, $a$) cell of the grid, $10^3$ planets were assigned with randomly drawn $\Teff$, semi-major axis and orbital phase. The orbital inclination $i$ was assumed to be the same as the disk inclination (see Table~\ref{tab:target_list}). 
If no inclination has been measured for the disk (see Table~\ref{tab:target_list}), $\sin(i)$ was randomly drawn with values uniformly distributed between (0,1). Eccentricities are assumed to be $e=0$. The other parameters were drawn from uniform distributions. Once planet orbits are simulated, their projected orbital separation is estimated, and if at that separation its effective temperature lies above the 1-D $\Teff$ limit curve (Sect.~\ref{sec:teff_limit}), they are considered as detected. Conversely, when they lie below the limits, they are considered as nondetectable\footnote{We note that this is a simplification, as contrast curves are not a fixed threshold \citep{Jensen-Clem2018}. However, as we are only interested in the average over the whole survey, this effect is neglected here.}. If the projected separation is larger than the image FoV as described in Sect.~\ref{sec:disks} or if the planet is located in the region of the image whose noise is dominated by disk signal and in which no statistically robust limits could have been calculated (Sect.~\ref{sec:contrast_curves}), we consider the planet as nondetected. The fraction of detected planets for each bin in the $\Teff-a$ parameter space provides then an estimate of the fraction of planets potentially detectable around each of our targets. 

This procedure provides three detection probability maps for each observed star (one for each assumed $\Rp$). The individual maps were then summed to generate a total completeness map of the survey, shown in Fig.~\ref{fig:prob_map} for $\Rp=2,3,5~\RJ$. For each combination of semi-major axis and effective temperature, these maps provide the number of stars to which the survey is complete. With a black marker we overplotted in the three maps (independently from its radius) the substellar companion HD101412\,B, the only one in the detection range of our survey. We note that since HD101412\,C is expected to be a stellar companion, its temperature is above the $\Teff$ range used in this work. Because of the bright disk ring detected in scattered light (Fig.~\ref{fig:disks}), statistically meaningful detection limits at the separation of PDS70 b and c could not be estimated. Not being in the investigated search space of the survey, the two protoplanets were not included in the main analysis of the demographic of protoplanets, even though in the next Sect.~\ref{sec:occurrence} we discuss the impact they would have on the results.

\section{Discussion}
\label{sec:discussion}

\subsection{Occurrence rate of forming planets}
\label{sec:occurrence}

\begin{figure}[b!]
    \includegraphics[width=\hsize]{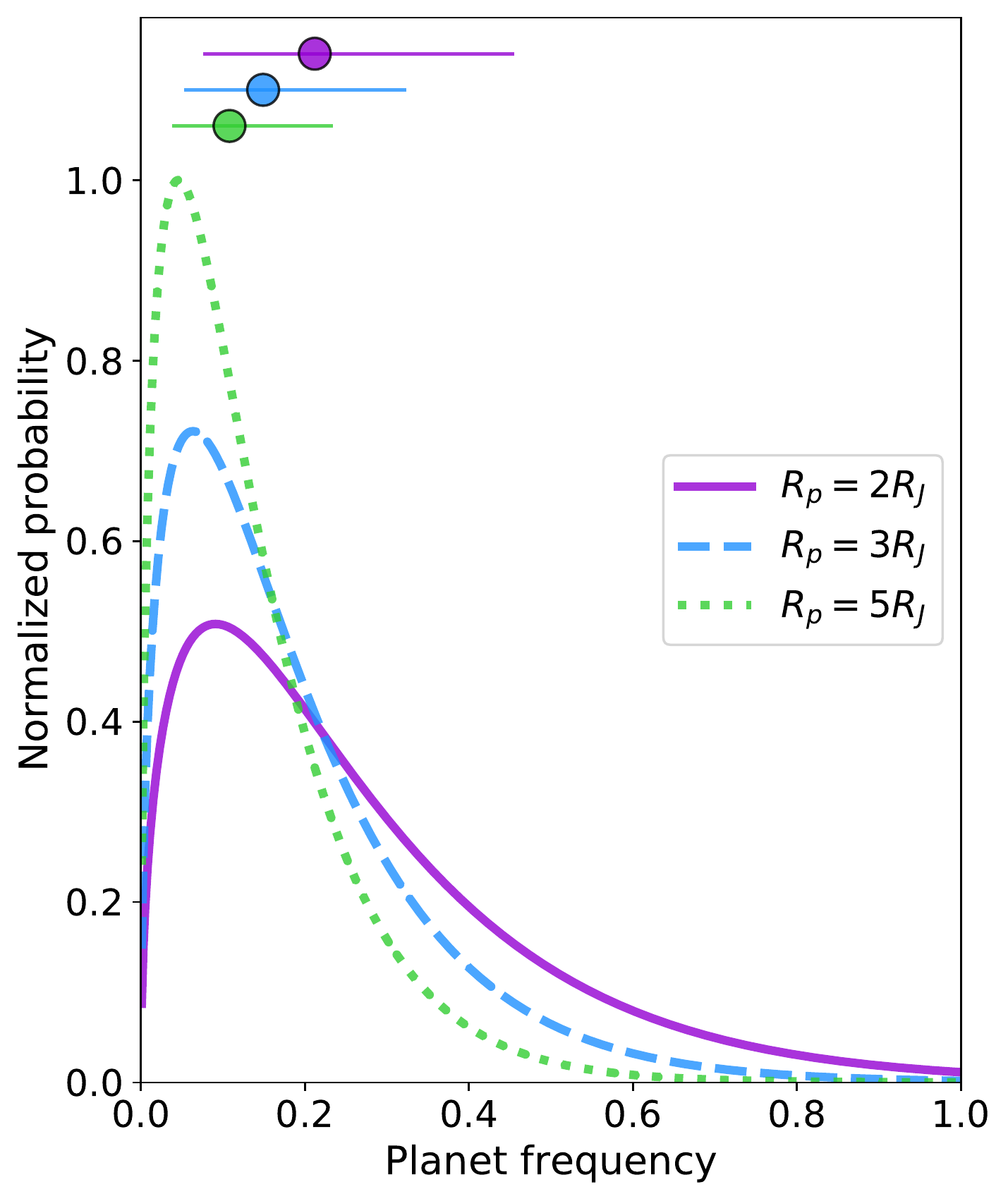}
    \caption{Posterior probability of the frequency of forming systems (20–500 au, 600–3000 K) for the different photometric radii $\Rp = 2,3,5~\RJ$. Filled circles represent the median of each distribution, while the corresponding errorbars give the 1$\sigma$ confidence intervals.}
    \label{fig:occurrence}
\end{figure}

We focused on the occurrence rate of forming gas giant planets with temperatures in the range $600-3000$~K orbiting with a semi-major axis in the range $20-500$~au, as these boundaries reflect the region of parameter space we are interested in and include all the protoplanets known to date. Some past works used population synthesis models to describe the underlying distribution of the planet demographic \citep[e.g., ][]{Vigan2017, Vigan2021}, relying on a set of assumptions on the disk and stellar properties as well as on planet dynamical evolution (or the lack thereof). Since little is empirically known about the distribution of forming planets, and many open questions remain on the disk properties, we undertook a simpler approach and assumed that planets are uniformly distributed in semi-major axis and temperature. We then integrate the completeness maps over the range mentioned above, obtaining a completeness to giant planets of 5.5, 7.8 and 10.8 targets for $\Rp = 2,3,5~\RJ$. Following \cite{Nielsen2019}, we considered a Poisson likelihood $L$ and a Jeffreys prior on the rate parameter of a Poisson distribution
\begin{equation}
L = \frac{\exp^{-\lambda}\lambda^{-k}}{k!}, \quad P(\lambda)=\sqrt{\frac{1}{\lambda}}
\label{eq:occurrence}
\end{equation}
where $\lambda$ is the expected number of planetary systems, that is the frequency multiplied by the completeness of our survey, and $k$ is the number of detected planetary systems ($k=1$ in our case). The probability of a given frequency to describe the population of detected planetary systems is plotted in Fig.~\ref{fig:occurrence}. From this distribution, it follows that the occurrence rate of forming giant planets with the characteristics described above is $21.2^{+24.3}_{-13.6}$\%, $14.8^{+17.5}_{-9.6}$\%, $10.8^{+12.6}_{-7.0}$\% assuming $\Rp = 2,3,5~\RJ$ (68\% confidence interval). 

Because of the known existence of the PDS70b and c protoplanets (which are located in a disk region excluded in this survey), we also estimated the occurrence rate of forming gas giants assuming two detected planetary systems ($k=2$ in Eq.~\ref{eq:occurrence}). We found values of $37.9^{+27.9}_{-19.6}$\%, $27.3^{+22.1}_{-14.3}$\% and $19.8^{+16.2}_{-10.4}$\% for $\Rp = 2,3,5~\RJ$. Even though these values are clearly larger than the nominal case presented above, they always fall within 1$\sigma$ from each other.

The choice of using effective temperatures to describe the planet population contrasts with what is usually done in other high-contrast imaging survey and makes a direct comparison of the detection rates rather difficult. However, if we assume that our temperature range corresponds to the mass range usually considered in high-contrast imaging surveys \citep[e.g.,][]{Nielsen2019, Vigan2021}, our occurrence rates generally agree with their findings: \citealt{Nielsen2019} found an occurrence rate of $24^{+13}_{-10}\%$ with a sample of 123 stars with $M_*>1.5~\Msun$, while \citealt{Vigan2021} found occurrence rates of $23.0^{+13.5}_{-9.7}$\% and $5.8^{+4.7}_{-2.8}$\% around BA and FGK stars respectively as part of the analysis of the first 150 targets of the SHINE sample.

\subsection{Sensitivity to distance and extinction from disk material}
\label{sec:dist_ext}
Figures~\ref{fig:map_distance} and \ref{fig:map_extinction} compare the detection probability maps presented in Fig.~\ref{fig:prob_map} with those obtained considering two different cases.
First, we limited the considered sample to targets with $d<150$~pc (Fig~\ref{fig:map_distance}), thus focusing on the nearest 23 targets. The survey completeness for this subsample is almost unchanged. The reason is that for targets at large distance from Earth the innermost region of the protoplanetary disks could not be investigated with NaCo, and at the same time only very hot objects could be detected as the flux scales with $1/d^2$. Figure~\ref{fig:map_distance} suggests that to constrain planet formation future surveys should focus on the nearby targets, as the innermost region of the disk could be investigated. Here we note the advantage provided by the new class of 30-40 meter telescopes, as the spatial resolution will be improved by a factor $\sim4-5$.

Second, we assume that circumstellar disk material absorbs and scatters light emitted by the planet reducing the flux escaping the circumstellar environment by $A_{L'}=1.0$~mag. This is a first order approximation, as different regions of the disk have different dust and gas surface densities and therefore planet flux is affected in very different ways. Furthermore, the presence of gaps and cavities as well as geometrical effects could dramatically change the extinction along the line of sight at different locations. As an example, \cite{Sanchis2020} employed hydrodynamical simulations to estimate the extinction in protoplanetary disks in 8 different bands and found that in the $L'$-band, a $2~\MJ$ planet opens a relatively small gap and suffers from an extinction of $\sim1.85$~mag at 50~au. To first order, we can expect that in disk regions without gaps the extinction is likely higher, and that in general it decreases with increasing separations as the surface densities drop (especially for the dust), allowing for a decreasing extinction factor. Additionally, more edge-on disks are generally expected to cause stronger attenuation, as the planet flux has to travel through a larger amount of disk material. However, modeling the disk extinction for each of our targets is beyond the scope of this paper, and here we just want to provide a sense of how this aspect may influence results. 

The maps for the extinction-corrected case are shown in Fig.~\ref{fig:map_extinction}, where they are compared to the standard case of Fig.~\ref{fig:prob_map}. We witness a strong decrease in sensitivity under the assumption that $A_{L'}=1.0$~mag and we are sensitive to objects several hundreds of kelvin hotter than in the nominal case presented in Sect.~\ref{sec:completeness}. Under the assumption of $A_{L'}=1.0$~mag, the occurrence rates as derived in Sect.~\ref{sec:occurrence} are $34.4^{+33.0}_{-21.8}$, $22.4^{+25.5}_{-14.4}$ and $14.4^{+16.8}_{-9.3}$ for $\Rp = 2,3,5~\RJ$ (68\% confidence interval), highlighting the strong impact of disk attenuation when trying to statistically constrain the population of protoplanets. Including a proper treatment of the dust and gas material in the disk would most likely increase extinction effects at small separations where pebbles are expected to be present close to the midplane. This would cause the effect of extinction to be even more dominant than in Fig.~\ref{fig:map_extinction}, strongly impacting the survey completeness. Furthermore, this analysis highlights how detection limits calculated in previous studies might be strongly affected by extincting circumstellar material, especially at short wavelengths \citep[$H$ and $K$ bands, where SPHERE and GPI operate, e.g.,][]{Asensio-Torres2021, Ginski2021, Mesa2019_HD163296}. Indeed, following the extinction law from \cite{Mathis1990}, an extinction of 1.0~mag in the $L'$ band corresponds to $A_K=2.0$~mag, $A_H=3.4$~mag and $A_V=14.6$~mag at shorter wavelengths. Hence, limits from those instruments for protoplanetary disks should be interpreted as lower bounds of our detection capabilities in protoplanetary disks. 

Figure~\ref{fig:map_extinction} calls for a better characterization of disk extinction properties, and how they may change with disk structures and as a function of the separation from the star and dust grain properties. This is going to be a crucial step in order to properly interpret current and future high-contrast imaging nondetections. In this context, observing in the MIR with upcoming instruments such as JWST/MIRI \citep{Rieke2015} or the ELT/METIS $N$-band filter \citep{Quanz2015_METIS} might overcome the obstacle of the circumstellar material impacting the intrinsic planet flux. For comparison, the extinction expected at $12~\mu$m for the case presented above is $A_{12~\mu\mathrm{m}}=0.5$~mag.

\subsection{Emission from accreting circumplanetary disks}
\label{sec:CPD}

\begin{figure}[t!]
    \includegraphics[width=\hsize]{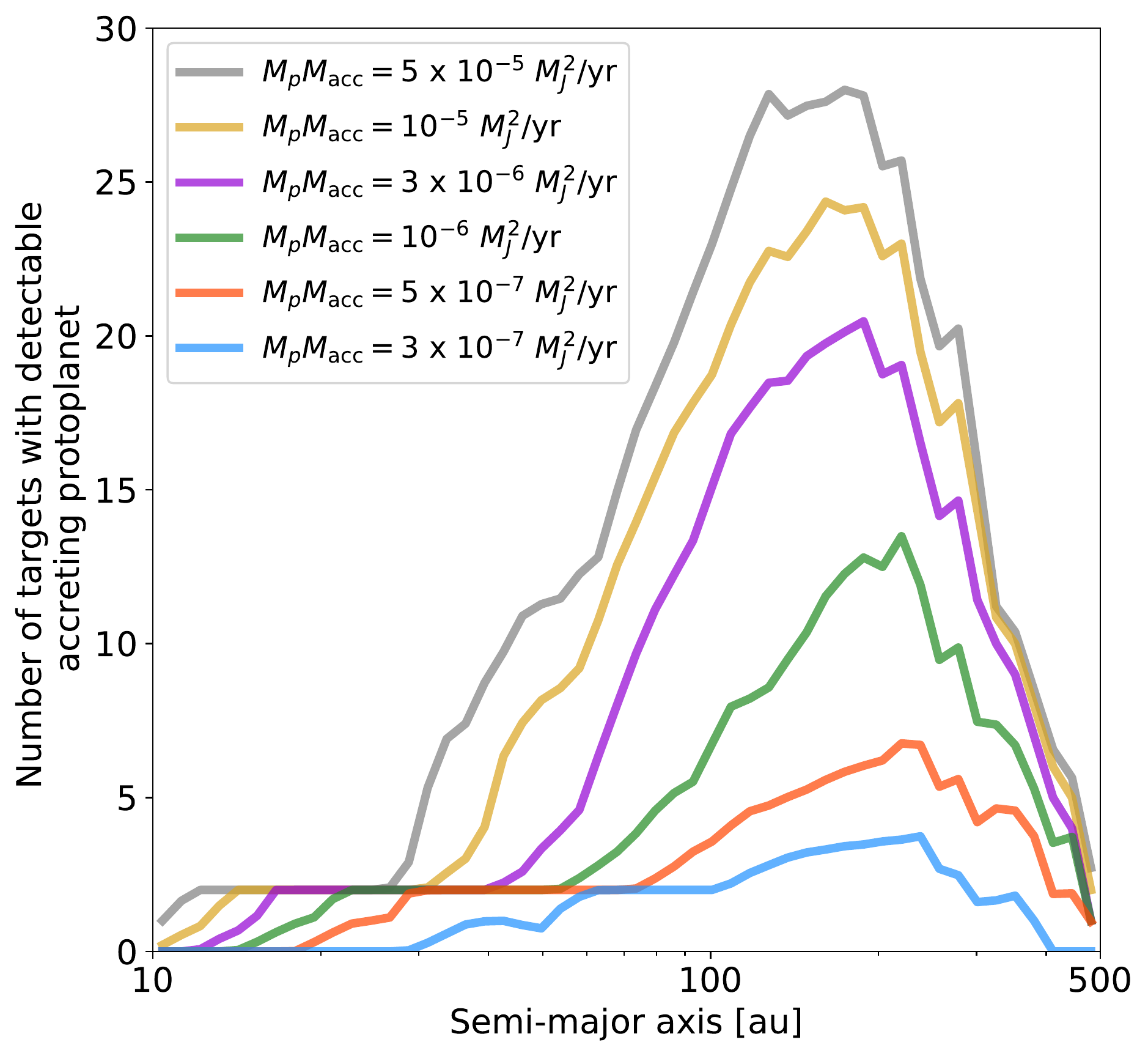}
    \caption{Depth of search for the NaCo-ISPY PPD sample, showing the number of stars to which the survey is complete for accreting objects as a function of $\Mp\Macc$ and semimajor axis. Massive planets and/or objects accreting at a high rate would have a higher chance to be detected and could be better constrained by our NaCo data. }
    \label{fig:accretion}
\end{figure}

Depending on the system properties, circumplanetary disk emission could be one order of magnitude brighter than the planet's photospheric emission \citep{Szulagyi2019}. In this section, we investigate the extreme case in which CPD emission dominates the protoplanet radiation, thus constraining ongoing accretion processes. We assumed that the accretion shock is fully thermalized, meaning that its emission can be described once again by a blackbody. Following \cite{Gullbring1998}, the accretion luminosity produced can be approximated by 
\begin{equation}
\Lacc = \frac{G~\Mp~\Macc}{\Rp}\left(1-\frac{\Rp}{R_\mathrm{in}} \right) \approx 1.25 \frac{G~\Mp~\Macc}{\Rp}
\label{eq:accretion}
\end{equation}
where $R_\mathrm{in}$ is the inner truncation radius of the CPD and $\Macc$ is the mass accretion rate onto the planet. The last step assumes that $R_\mathrm{in}=5~\Rp$ \citep[e.g., ][]{Cugno2019_a}. In this section we assume $\Rp=3~\RJ$, but the same reasoning can be applied to the other radii. Once the accretion shock radius has been fixed, $\Lacc$ only depends on the $\Mp\Macc$ term, for which we considered different values as reported in Fig.~\ref{fig:accretion}. For each $\Mp\Macc$, we estimated the temperature able to produce a blackbody emission with bolometric luminosity equal to the total accretion luminosity from Eq.~\ref{eq:accretion}. We then estimated for how many targets each accretion scenario would have been detected based on the central map of Fig.~\ref{fig:prob_map}. 

Figure~\ref{fig:accretion} clearly indicates that more massive planets and/or objects accreting at a higher rate have a higher chance to be detected. These results suggest that high-mass planets accreting at high rates are rare, especially at separations larger than $100$~au, confirming previous searches for accreting protoplanets \citep{Huelamo2018, Cugno2019_a, Zurlo2020, Xie2020}.

\section{Summary and outlook}
\label{sec:conclusions}

We have presented the NaCo-ISPY protoplanetary disk sample, which included observations of 45 young stars with the NaCo instrument at the VLT, searching for forming planets. Observations were carried out in the $L'$-band and were meant to obtain the highest sensitivity to faint companions. Although the sample size is relatively small compared to other large imaging surveys, which observed hundreds of stars \citep[e.g., GPIES and SHINE,][]{Nielsen2019, Vigan2021}, the uniqueness of this work lies in the presence of protoplanetary disks surrounding each of the targets. Many of the investigated disks show signposts of ongoing planet formation, such as substructures in the dust and gas distribution likely due to the interaction with embedded forming planets. The PPD sample of the NaCo-ISPY survey offers therefore a unique possibility to probe the population of young forming planets around some of the nearest disks, allowing us to learn important lessons for future surveys. 

\begin{enumerate}
    \item We detected 15 companions around 13 targets (out of 45), 2 of which with planetary masses (PDS70\,b and c), and at least one with estimated mass in the brown dwarf regime (HD101412\,C, see also \citealt{Rich2022} and Ruh et al., in prep.).
    \item Disk signals were detected around 17 targets. For two of them, HD72106 and T\,CrA, this is the first spatially resolved image of the outer disk.
    \item We showed that the presence of disk signals in the final residuals breaks the assumption of Gaussian noise, therefore strongly biasing results in those regions which should not be trusted. 
    \item We highlighted the strong dependence of the mass-to-luminosity conversion for forming planets on the underlying assumptions, showing the difficulties in determining planet masses and how uncertainties dominate every possible result, especially in the $L'$ band. To overcome these obstacles, we propose a new approach, relying on the study of the SEDs for the forming planets PDS70\,b and c. 
    \item We estimated the occurrence of forming companions with temperatures in the range $600-3000$ K with semi-major axes in the range $20-500$ au to be $21.2^{+24.3}_{-13.6}$\%, $14.8^{+17.5}_{-9.6}$\%, $10.8^{+12.6}_{-7.0}$\% for blackbodies with $\Rp = 2,3,5~\RJ$ respectively.
    \item We show that extinction might be a key factor in the low detection rates, and more advanced calculations of its effect on the flux observed from protoplanets are warranted in order to fully understand and quantify its impact as a function of the separation from the star. 
\end{enumerate}

With its MIR capabilities and sensitivity, the James Webb Space Telescope might be able to detect embedded forming companions, increasing the number of directly imaged protoplanets and proving the connection between these objects and the disk features observed with ALMA and high-contrast imagers.


\begin{acknowledgements}
We would like to thank the anonymous referee, whose careful and constructive comments improved the quality of this manuscript.
GC thanks the Swiss National Science Foundation for financial support under grant number P500PT\_206785.
GC and SPQ thank the Swiss National Science Foundation for financial support under grant number 200021\_169131.
TDP is supported by Deutsche Forschungsgemeinschaft (DFG) grants Kr 2164/14-2 and Kr 2164/15-2.
T. H. acknowledges support from the European Research Council under the Horizon 2020 Framework Program via the ERC Advanced Grant Origins 832428.
GMK is supported by the Royal Society as a Royal Society University Research Fellow.
Part of this work has been carried out within the frame of the National Center for Competence in Research PlanetS supported by the Swiss National Science Foundation (SNSF). SPQ acknowledges the financial support of the SNSF.

This work has made use of data from the European Space Agency (ESA) mission {\it Gaia} (\url{https://www.cosmos.esa.int/gaia}), processed by the {\it Gaia} Data Processing and Analysis Consortium (DPAC, \url{https://www.cosmos.esa.int/web/gaia/dpac/consortium}). Funding for the DPAC has been provided by national institutions, in particular the institutions participating in the {\it Gaia} Multilateral Agreement. 

\end{acknowledgements}

%
%

\bibliographystyle{aa}
\bibliography{ISPY_PPD.bib}

\begin{appendix} 
\section{Additional tables}

\longtab[1]{
\begin{landscape}
\def\arraystretch{1.45}
\small
\begin{longtable}{lllllllllllll}
\caption{\label{tab:target_list}Target sample and their main properties used in this work.}\\
\hline
\hline
Target\tablefootmark{a} & RA\tablefootmark{b} & DEC\tablefootmark{b}    & d [pc]\tablefootmark{c} & $L'$ [mag]       &  Age [Myr]\tablefootmark{d} & $M_*$ $[M_\odot]$ & Sp. Type  &$\Teff$ [K]& Incl. [$^\circ$] & $\Rd$ [au]\tablefootmark{e} & $\Rg$ [au]\tablefootmark{e} & Refs      \\ 
\midrule
\endfirsthead
\caption{continued}\\
\hline
Target\tablefootmark{a} & RA\tablefootmark{b} & DEC\tablefootmark{b}    & d [pc]\tablefootmark{c} & $L'$ [mag]       &  Age [Myr]\tablefootmark{d} & $M_*$ $[M_\odot]$ & Sp. Type  &$\Teff$ [K]& Incl. [$^\circ$] & $\Rd$ [au]\tablefootmark{e} & $\Rg$ [au]\tablefootmark{e} & Refs      \\ 
\midrule
\endhead
\hline
\endfoot
\hline
\endlastfoot
V892\,Tau& 04:18:40.61   & 28:19:15.64   & $134.5 \pm 1.5$    & $4.80 \pm 0.19$  & 5.0            & $2.8$   & A0  & 9550      & 54  & 60  & 240   & 1,18,33,35\\
HD 283571  & 04:21:57.41   & 28:26:35.53   & $138.2 \pm 3.9$    & $3.70 \pm 0.50$  & 4.5            & $2.9$   & F7  & 6220      & 65  & 65  & ...   & 2,20,40   \\
HL\,Tau  & 04:31:38.43   & 18:13:57.65   & 140.0              & $4.64 \pm 0.35$  & 1.5            & $1.2$   & K5  & 4400      & 47  & 100 & ...   & 19,31     \\
HD 31648   & 04:58:46.26   & 29:50:36.99   & $156.2 \pm 1.3$    & $4.42 \pm 0.25$  & 7.0            & $2.3$   & A5  & 8250      & 36  & 102 & 333   & 2,6,39,40 \\
HD 34282   & 05:16:00.47   & $-$09:48:35.39  & $308.6 \pm 2.2$    & $6.60 \pm 0.10$  & $6.4\pm2.6$    & $1.8$   & A3  & 9400      & 59  & 354 & 956   & 2,4,24    \\
HD 35187   & 05:24:01.17   & 24:57:37.57   & $161.6 \pm 1.3$    & $5.00 \pm 0.18$  & $9.0\pm2.0$    & $1.8$   & A2  & 8900      & ... & ... & ...   & 2,7       \\
HD 36112   & 05:30:27.52   & 25:19:57.08   & $155.9 \pm 0.8$    & $4.40 \pm 0.27$  & $8.9\pm2.0$    & $1.9$   & A8  & 8200      & 21  & 99  & 154   & 2,13,11,17\\
HD 36910   & 05:35:58.46   & 24:44:54.08   & $149.4 \pm 1.3$    & $5.00 \pm 0.22$  & $9.9\pm2.9$    & $1.0$   & F5  & 6800      & 35  & 75  & 160   & 2,13,30   \\
HD 37411   & 05:38:14.50   & $-$05:25:13.31  & $350.4 \pm 5.9$    & $6.28 \pm 0.10$  & $9.0\pm4.5$    & $1.9$   & A0  & 9750      & ... & ... & ...   & 3         \\
HD 37806   & 05:41:02.29   & $-$02:43:00.72  & $401.6 \pm 4.4$    & $4.20 \pm 0.30$  & $1.6\pm0.6$    & $3.5$   & B9  & 10475     & ... & ... & ...   & 2,8       \\
HD 38120   & 05:43:11.89   & $-$04:59:49.88  & $384.5 \pm 5.0$    & $6.29 \pm 0.10$  & $3.5\pm1.4$    & $2.6$   & B9  & 10700     & ... & ... & ...   & 2,8       \\
HD 259431  & 06:33:05.19   & 10:19:19.98   & $653.4 \pm 11.6$   & $4.00 \pm 0.35$  & $0.4\pm0.3$    & $4.0$   & B6  & 14000     & ... & ... & ...   & 2,8       \\
NX\,Pup     & 07:19:28.28   & $-$44:35:11.23  & $410.0\pm82.0$     & $4.20 \pm 0.40$  & $4.9\pm2.4$    & $1.9$   & A1  & 7000      & ... & ... & ...   & 3,8       \\  
HD 58647   & 07:25:56.10   & $-$14:10:43.55  & $304.6 \pm 2.4$    & $3.90 \pm 0.40$  & $0.4\pm0.1$    & $3.3$   & B9  & 10500     & ... & ... & ...   & 2,43      \\
HD 72106   & 08:29:34.90   & $-$38:36:21.13  & $382.5 \pm 11.3$   & $6.79 \pm 0.04$  & $3.8\pm1.5$    & $1.8$   & A0  & 9500      & 51\tablefootmark{i}& ... & ... & 3,37 \\
HD 85567   & 09:50:28.54   & $-$60:58:02.95  & $1047.4 \pm 18.0$  & $4.27 \pm 0.34$  & $0.3\pm0.2$    & $4.9$   & B8  & 13000     & ... & ... & ...   & 2,3       \\
TW\,Hya     & 11:01:51.90   & $-$34:42:17.03  & $60.1 \pm 0.1$     & $6.97 \pm 0.10$  & $6.3\pm3.7$    & $0.8$   & K6  & 4000      & 7   & 59  & 184   & 2,13,34   \\ 
HD 95881   & 11:01:57.62   & $-$71:30:48.31  & $1109.9 \pm 24.3$  & $4.16 \pm 0.25$  & $0.3\pm0.1$    & $5.5$   & A0  & 10000     & ... & ... & ...   & 8         \\
HD 97048   & 11:08:03.31   & $-$77:39:17.49  & $184.4 \pm 0.8$    & $4.50 \pm 0.30$  & $3.9\pm1.7$    & $2.2$   & A0  & 10400     & 40  & 185 & 602   & 2,13,21,47\\
HD 98922   & 11:22:31.67   & $-$53:22:11.46  & $650.9 \pm 8.8$    & $2.20 \pm 0.50$  & $0.8\pm0.4$    & $5.4$   & B9  & 10500     & ... & ... & ...   & 2,3       \\
HD 100453  & 11:33:05.58   & $-$54:19:28.54  & $103.8 \pm 0.2$    & $4.36 \pm 0.35$  & $6.5\pm4.8$    & $1.7$   & A9  & 7250      & 29  & 41  & 113   & 2,8,25    \\
HD 100546  & 11:33:25.44   & $-$70:11:41.24  & $108.1 \pm 0.4$    & $3.95 \pm 0.27$  & $5.0\pm1.3$    & $1.9$   & A0  & 9800      & 42  & 50  & 389   & 2,13,14,27\\
HD 101412  & 11:39:44.46   & $-$60:10:27.72  & $412.2 \pm 2.5$    & $5.75 \pm 0.10$  & $6.4\pm1.8$    & $2.0$   & A3  & 9750      & ... & ... & ...   & 3         \\
HD 104237  & 12:00:05.09   & $-$78:11:34.56  & $106.6 \pm 0.5$    & $3.48 \pm 0.45$  & $4.9\pm1.9$    & $2.3$   & A0  & 8000      & 18  & ... & ...   & 2,3,44    \\
PDS70     & 14:08:10.15   & $-$41:23:52.58  & $112.4 \pm 0.2$    & $7.93 \pm 0.03$  & $8.0\pm1.0$    & $0.9$   & K7  & 4100      & 52  & 100 & 224   & 9,26,29,38\\
HD 139614  & 15:40:46.38   & $-$42:29:53.54  & $133.6 \pm 0.5$    & $5.67 \pm 0.11$  & $11.8\pm3.6$   & $1.6$   & A9  & 7800      & 18  & ... & 200   & 4,36      \\
HD 141569  & 15:49:57.75   & $-$03:55:16.34  & $111.6 \pm 0.4$    & $6.07 \pm 0.10$  & $9.0\pm4.5$    & $2.0$   & A2  & 9750      & 57  & 222 & 279   & 2,3,28    \\
V* IM Lup  & 15:56:09.18   & $-$37:56:06.12  & $155.8 \pm 0.5$    & $6.82 \pm 0.10$  & $1.7\pm1.2$    & $0.4$   & M0  & 4400      & 47  & 260 & 493   & 2,4,23,39 \\   
HD 142666  & 15:56:40.02   & $-$22:01:40.00  & $146.3 \pm 0.5$    & $4.94 \pm 0.24$  & $7.1\pm0.3$    & $1.6$   & A8  & 7600      & 62  & 53  & 171   & 34        \\
HD 142527  & 15:56:41.89   & $-$42:19:23.24  & $159.3 \pm 0.7$    & $3.97 \pm 0.39$  & $6.2\pm1.5$    & $2.2$   & F6  & 6400      & 38  & 306 & 408   & 2,4,15,36 \\
HD 143006  & 15:58:36.91   & $-$22:57:15.22  & $167.3 \pm 0.5$    & $5.63 \pm 0.13$  & $6.6\pm0.3$    & $1.8$   & G7  & 5600      & 19  & 78  & 154   & 34        \\
MY\,Lup  & 16:00:44.53   & $-$41:55:31.00  & $157.2 \pm 0.9$    & $7.97 \pm 0.03$  & $6.3$          & $1.2$   & K0  & 5100      & 73  & 77  & 192   & 34        \\  
2MASS J1604\,\tablefootmark{f}&16:04:21.65&-21:30:28.55&$145.3\pm0.6$&$7.45\pm0.05$& $11.1\pm3.3$   & $1.0$   & K2  & 4600      & 10  & 79  & 400   & 13,41,42  \\
HD 144432  & 16:06:57.95   & $-$27:43:09.76  & $154.8 \pm 0.6$    & $5.10 \pm 0.21$  & $8.9\pm1.5$    & $2.0$   & A9  & 7400      & ... & ... & ...   & 2,4       \\
TYC\,7851\tablefootmark{g}&16:08:30.70&$-$38:28:26.85& $153.4\pm0.7$ & $8.22\pm0.04$  & $8.3\pm4.4$    & $1.4$   & K2  & 4800      & 74  & 140 & 302   & 4,22      \\
HD 144668  & 16:08:34.29   & $-$39:06:18.34  & $158.6 \pm 0.9$    & $2.50 \pm 0.50$  & $4.0\pm0.4$    & $2.2$   & A7  & 8400      & ... & 42  & ...   & 2,4,32    \\
HD 145263  & 16:10:55.11   & $-$25:31:21.67  & $141.9 \pm 0.4$    & $7.74 \pm 0.03$  & $11.0$         & $1.5$   & F0  & 7250      & ... & ... & ...   & 2,10      \\
Elias\,2$-$27 & 16:26:45.03   & $-$24:23:07.81  & $110.1 \pm 10.3$   & $7.24 \pm 0.04$  & $0.8$          & $0.5$   & M0  & 3900      & 56  & 240 & ...   & 16,23     \\
WLY 2$-$48   & 16:27:37.19   & $-$24:30:35.03  & $136.3 \pm 1.9$    & $5.43 \pm 0.15$  & $6.9\pm2.2$    & $2.2$   & B5  & 9000      & 50  & ... & 180   & 4,48      \\
AS\,209  & 16:49:15.30   & $-$14:22:08.64  & $121.2 \pm 0.4$    & $6.22 \pm 0.08$  & $6.0\pm0.4$    & $2.8$   & K5  & 4300      & 35  & 127 & 280   & 2,34      \\
HD 152404  & 16:54:44.85   & $-$36:53:18.56  & $139.8 \pm 0.6$    & $5.34 \pm 0.16$  & $>12.0$        & $1.4$   & F5  & 6200      & 71  & 53  & ...   & 4,35      \\
KK\,Oph  & 17:10:08.12   & $-$27:15:18.80  & $167.0 \pm 3.7$    & $4.13 \pm 0.31$  & $15.2\pm7.6$   & $1.6$   & A6  & 8500      & ... & ... & ...   & 3         \\
HD 158643  & 17:31:24.95   & $-$23:57:45.52  & $125.7 \pm 1.7$    & $3.30 \pm 0.40$  & $1.2\pm0.6$    & $3.1$   & A0  & 9800      & ... & ... & ...   & 2,8       \\
HD 163296  & 17:56:21.29   & $-$21:57:21.88  & $101.0 \pm 0.4$    & $3.46 \pm 0.44$  & $7.1\pm0.6$    & $2.1$   & A1  & 9300      & 47  & 137 & 478   & 2,34      \\
HD 319139  & 18:14:10.48   & $-$32:47:34.52  & $71.5 \pm 0.1$     & $7.13 \pm 0.10$  & $12.9\pm5.7$   & $1.1$   & K6  & 4000      & 35  & ... & 306   & 4,21      \\
HD 169142  & 18:24:29.78   & $-$29:46:49.33  & $114.9 \pm 0.4$    & $6.00 \pm 0.10$  & $12.3\pm6.4$   & $1.5$   & F1  & 7400      & 13  & 83  & 178   & 13,46     \\
V* R\,CrA   & 19:01:53.68   & $-$36:57:08.14  & $125.2 \pm 7.6$    & $1.00 \pm 0.40$  & $1.5\pm1.2$    & $3.3$   & B5  & 8150      & ... & ... & ...   & 8,12      \\   
V* T\,CrA   & 19:01:58.79   & $-$36:57:50.33  & 130                & $6.20 \pm 0.20$  & 22.8           & 1.5     & F0  & 7250      & 77\tablefootmark{i}& ...  & ... & 5         \\  
HD 179218  & 19:11:11.25   & 15:47:15.63   & $260.1 \pm 2.2$    & $4.52 \pm 0.25$  & $1.9\pm0.9$    & $2.7$   & A0  & 9600      & ... & ... & 160\tablefootmark{h}& 2,4,45  \\
HD 190073  & 20:03:02.51   & 05:44:16.66   & $847.9 \pm 22.5$   & $4.38 \pm 0.31$  & $0.2\pm0.1$   & $4.7$    & A2  & 9500      & ... & ... & ...   & 2,8       \\

\end{longtable}
\tablefoot{Uncertainties are not given when not reported in the reference. 
\tablefoottext{a}{If available, we use the HD number as main source ID. }
\tablefoottext{b}{ICRS, from {\it Gaia} DR3 \citep{Gaia2022} where available (Epoch 2016.0). }
\tablefoottext{c}{Distances and their uncertainties are inferred from {\it Gaia} DR3 parallaxes \citep{Gaia2022}, except for HL Tau, NX\,Pup and T\,CrA, which are taken from \cite{ALMA2015}, \cite{Fairlamb2015} and \cite{Manoj2006}, respectively. } 
\tablefoottext{d}{Ages compiled here are taken from the literature and derived with different methods and different treatment of uncertainties. Some references only summarize different other age estimation attempts. }
\tablefoottext{e}{Outer disk radii are compiled from the literature and originate from different methods. Only outer disk radii directly imaged or inferred with ALMA were considered. They are corrected for new {\it Gaia} DR3 distances when necessary. }
\tablefoottext{f}{2MASS J16042165$-$2130284. }
\tablefoottext{g}{TYC 7851$-$810$-$1. }
\tablefootmark{h}{Based on scattered light images tracing small dust particles expected to couple well with the gas.}
\tablefootmark{i}{Inclination estimated in this work in Sect.~\ref{sec:disk_det}.}}
\tablebib{(1) \cite{Hamidouche2010}; 
(2) \cite{Kervella2019}; 
(3) \cite{Fairlamb2015}; 
(4) \cite{Garufi2018}; 
(5) \cite{Manoj2006}; 
(6) \cite{Montesinos2009}; 
(7) \cite{Meeus2012}; 
(8) \cite{Vioque2018}; 
(9) \cite{Wang2021}; 
(10) \cite{Chen2014}; 
(11) \cite{Boehler2018};
(12) \cite{Cugno2019_b}; 
(13) \cite{Asensio-Torres2021}; 
(14) \cite{Pineda2019}; 
(15) \cite{Boehler2017}; 
(16) \cite{Andrews2018}; 
(17) \cite{Dong2018}; 
(18) \cite{Long2021}; 
(19) \cite{Skinner2020}; 
(20) \cite{Garufi2019}; 
(21) \cite{Law2022}; 
(22) \cite{Ansdell2018}; 
(23) \cite{Huang2018}; 
(24) \cite{vanderplas2017}; 
(25) \cite{vanderplas2019}; 
(26) \cite{Keppler2019}; 
(27) \cite{Pineda2014}; 
(28) \cite{DiFolco2020}; 
(29) \cite{Facchini2021}; 
(30) \cite{UbeiraGabellini2019}; 
(31) \cite{ALMA2015}; 
(32) \cite{Panic2021}; 
(33) \cite{Liu2011}; 
(34) \cite{Long2022}; 
(35) \cite{Stapper2022}; 
(36) \cite{Bohn2022}; 
(37) \cite{Schegerer2009}; 
(38) \cite{Benisty2021}; 
(39) \cite{Law2021_MAPS3}; 
(40) \cite{Long2018}; 
(41) \cite{Canovas2017}; 
(42) \cite{VanderMarel2015}; 
(43) \cite{Kurosawa2016}; 
(44) \cite{Grady2004}; 
(45) \cite{Garufi2022}; 
(46) \cite{Fedele2017}; 
(47) \cite{Pinte2019}; 
(48) \cite{Bruderer2014}.
}
\end{landscape}
}

\longtab[2]{
\begin{table*}[t!]
\caption{Observations of the NaCo-ISPY PPD targets.}
\small
\def\arraystretch{1.25}
\begin{tabular}{llllllllll}
Target          & Obs. date     & Seeing\tablefootmark{a}    & Airmass   & Field rot. & AGPM  & PSF stab.  & DIT\tablefootmark{b}   & ToT\tablefootmark{c}   & Comments\tablefootmark{e}  \\
                & [yyyy-mm-dd]  &[$^{\prime\prime}$]&(min/max)& [$^\circ$]&     & [\%]      & [s]   & [min]   &      \\\hline
V* V892 Tau     & 2016-12-10    &$1.9\pm0.5$& 1.7/2.3   & 50         & y     & 13.4      & 0.35  & 130.6      & (1)  \\
HD 283571       & 2016-12-09    &$1.3\pm0.4$& 1.7/2.0   & 42         & y     & 7.0       & 0.35  & 104.4      & (1)  \\
V* HL Tau       & 2018-10-19    &$0.7\pm0.1$& 1.4/1.7   & 45         & y     & 4.5       & 0.35  & 93.3       & (1)  \\ 
HD 31648        & 2017-11-01    &$0.8\pm0.1$& 1.7/2.3   & 66         & y     & 9.0       & 0.35  & 168.0      &      \\
HD 34282        & 2016-11-07    &$0.7\pm0.2$& 1.0/1.2   & 118        & y     & 4.6       & 0.25  & 135.4      &      \\
HD 35187        & 2019-01-15    &$0.7\pm0.2$& 1.5/1.9   & 56         & y     & 10.2      & 0.35  & 114.9      &      \\
HD 36112        & 2019-01-16    &$0.6\pm0.1$& 1.6/1.8   & 63         & y     & 2.0       & 0.35  & 131.8      & (1)  \\
HD 36910        & 2018-11-27    &$0.5\pm0.1$& 1.5/2.0   & 69         & y     & 1.9       & 0.35  & 175.6      &      \\
HD 37411        & 2017-11-02    &$0.6\pm0.1$& 1.1/1.1   & 69         & y     & 4.1       & 0.35  & 84.0       &      \\
HD 37806        & 2017-10-30    &$0.8\pm0.2$& 1.1/1.2   & 55         & y     & 5.1       & 0.35  & 78.8       &      \\
HD 38120        & 2017-10-29    &$0.5\pm0.1$& 1.1/1.1   & 66         & y     & 2.8       & 0.35  & 84.0       &      \\
V* NX Pup       & 2018-02-22    &$0.5\pm0.1$& 1.1/1.3   & 78         & y     & 2.1       & 0.3   & 130.0      &      \\    
HD 58647        & 2018-02-23    &$0.6\pm0.1$& 1.0/1.2   & 126        & y     & 4.1       & 0.3   & 123.5      &      \\
HD 72106        & 2016-12-12    &$0.6\pm0.1$& 1.0/1.2   & 130        & n     & 3.8       & 0.2   & 89.5       &   \\
V* TW Hya       & 2016-05-03    &$0.8\pm0.3$& 1.0/1.1   & 135        & y     & 4.1       & 0.35  & 106.2      & (4)  \\        
HD 97048        & 2016-05-02    &$0.6\pm0.1$& 1.7/1.8   & 68         & y     & 5.6       & 0.35  & 151.7      &   \\
HD 100453       & 2016-05-09    &$2.4\pm0.6$& 1.2/1.3   & 85         & y     & 13.4      & 0.25  & 97.5       &   \\
HD 100546       & 2017-03-15    &$0.8\pm0.3$& 1.4/1.5   & 64         & y     & 2.2       & 0.25  & 110.4      & (2)  \\
HD 101412       & 2017-03-17    &$0.7\pm0.2$& 1.2/1.3   & 74         & y     & 10.8      & 0.3   & 132.5      & (2)  \\
HD 104237       & 2017-05-16    &$1.0\pm0.2$& 1.7/2.0   & 52         & y     & 4.8       & 0.35  & 152.3      &   \\
PDS 70          & 2016-06-01    &$0.4\pm0.1$& 1.0/1.2   & 84         & n     & 5.3       & 0.2   & 61.7       &   \\
HD 139614       & 2017-05-01    &$0.6\pm0.3$& 1.1/1.2   & 102        & y     & 4.0       & 0.35  & 98.6       & (1)  \\
HD 141569       & 2019-04-13    &$0.6\pm0.1$& 1.1/1.2   & 107        & y     & 2.8       & 0.35  & 130.1      &   \\
V* IM Lup       & 2017-05-15    &$1.2\pm0.2$& 1.0/1.1   & 120        & n     & 5.9       & 0.2   & 78.5       &   \\ 
HD 142666       & 2019-05-20    &$1.1\pm0.4$& 1.0/1.0   & 149        & y     & 4.6       & 0.35  & 40.3       &   \\
HD 142527       & 2017-05-17    &$1.0\pm0.2$& 1.1/1.2   & 108        & y     & 8.0      & 0.35  & 129.5      &   \\
HD 143006       & 2019-04-12    &$0.9\pm0.2$& 1.0/1.2   & 162        & y     & 3.7       & 0.35  & 114.3      &   \\
V* MY Lup       & 2018-06-04    &$1.0\pm0.2$& 1.0/1.2   & 114        & n     & 7.1       & 0.2   & 100.8      &   \\ 
2MASS J1604     & 2016-05-31    &$0.8\pm0.1$& 1.0/1.1   & 147        & n     & 6.3       & 0.2   & 65.0       & (2), (3)\\
HD 144432       & 2019-05-19    &$1.0\pm0.1$& 1.0/1.1   & 165        & y     & 6.2       & 0.35  & 67.7       & (2)   \\
TYC 7851        & 2018-06-06    &$0.6\pm0.1$& 1.0/1.1   & 111        & n     & 3.4       & 0.2   & 76.8       &   \\
HD144668        & 2017-06-16    &$0.7\pm0.1$& 1.0/1.1   & 61         & y     & 18.6      & 0.35  & 53.7       & (2)  \\
HD 145263       & 2019-05-25    &$1.1\pm0.1$& 1.0/1.0   & 179        & n     & 3.2       & 0.2   & 45.4       & (5) \\
Elias 2-27      & 2019-04-14    &$0.4\pm0.1$& 1.0/1.2   & 192\tablefootmark{d}& n& 6.8   & 0.2   & 63.3       & (1) \\
WLY 2-48        & 2016-07-31    &$0.6\pm0.1$& 1.0/1.1   & 55         & y     & 4.5       & 0.35  & 63.0       & (2)   \\
EM* AS209       & 2019-07-20    &$0.8\pm0.2$& 1.0/1.3   & 135        & n     & 1.8       & 0.2   & 152.0      & (1)   \\
HD 152404       & 2019-05-18    &$0.8\pm0.1$& 1.0/1.2   & 95         & y     & 5.6       & 0.35  & 79.9       &   \\
V* KK Oph       & 2016-08-01    &$0.5\pm0.1$& 1.0/1.1   & 171        & y     & 5.1       & 0.35  & 98.6       &   \\ 
HD 158643       & 2019-06-24    &$0.8\pm0.1$& 1.0/1.0   & 168        & y     & 12.5      & 0.3   & 46.0       &   \\
HD 163296       & 2019-07-13    &$0.5\pm0.1$& 1.0/1.0   & 151\tablefootmark{d}& y& 3.0   & 0.35  & 48.4       &   \\
HD 319139       & 2016-05-03    &$0.5\pm0.2$& 1.0/1.1   & 161        & y     & 28.4       & 0.35  & 106.2      &   \\
HD 169142       & 2017-05-18    &$0.6\pm0.1$& 1.0/1.2   & 107        & y     & 3.7       & 0.35  & 109.7      &   \\
V* R\,CrA        & 2018-06-06    &$0.6\pm0.1$& 1.0/1.2   & 120        & y     & 22.8      & 0.1082& 47.5       &   \\   
V* T\,CrA        & 2017-05-15    &$1.2\pm0.2$& 1.0/1.1   & 122        & y     & 4.7       & 0.35  & 106.8      &   \\
HD 179218       & 2016-05-02    &$0.8\pm0.3$& 1.3/1.6   & 64         & y     & 3.0      & 0.35  & 126.1      &   
\end{tabular}
\tablefoot{\tablefoottext{a}{DIMM (Differential Image Motion Monitor) seeing.} \tablefoottext{b}{DIT = Detector integration time, i.e., exposure time per image frame. }\tablefoottext{c}{ToT = Time on Target. }\tablefoottext{d}{Passage through the sky region around zenith not accessible to the VLT telescopes.} \tablefoottext{e}{(1) Only unsaturated PSF frames taken at the beginning of the observing sequence were used; (2) Only unsaturated PSF frames taken at the end of the observing sequence were used; (3) The first 21 cubes were not used; (4) only 2 quadrants before and after the observing sequence were used; (5) one cube was removed from the unsaturated PSF sequence because the star was not well-centered in the quadrant.}\tablefoottext{f}{2MASS J16042165$-$2130284. }
\tablefoottext{g}{TYC 7851$-$810$-$1. }}
\label{tab:observations}
\vspace{2cm}
\end{table*}
}

\section{PSF-subtracted images}

\begin{figure*}
    \centering
    \includegraphics[width=0.87\hsize]{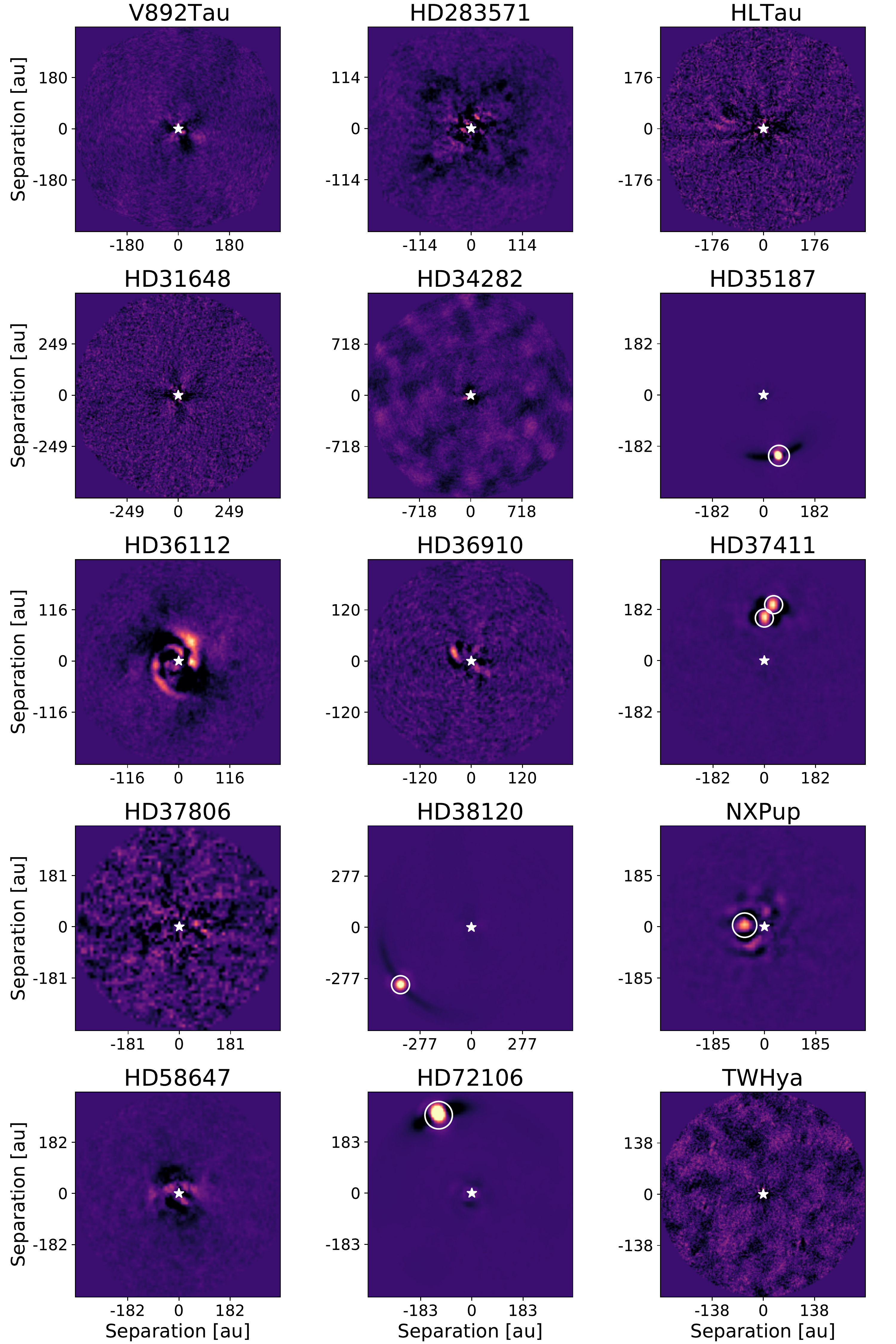}
    \caption{Final PCA-ADI residuals of the ISPY PPD sample. White circles indicate the position of companions.}
    \label{fig:residuals_1}
\end{figure*}

\begin{figure*}
    \centering
    \includegraphics[width=0.87\hsize]{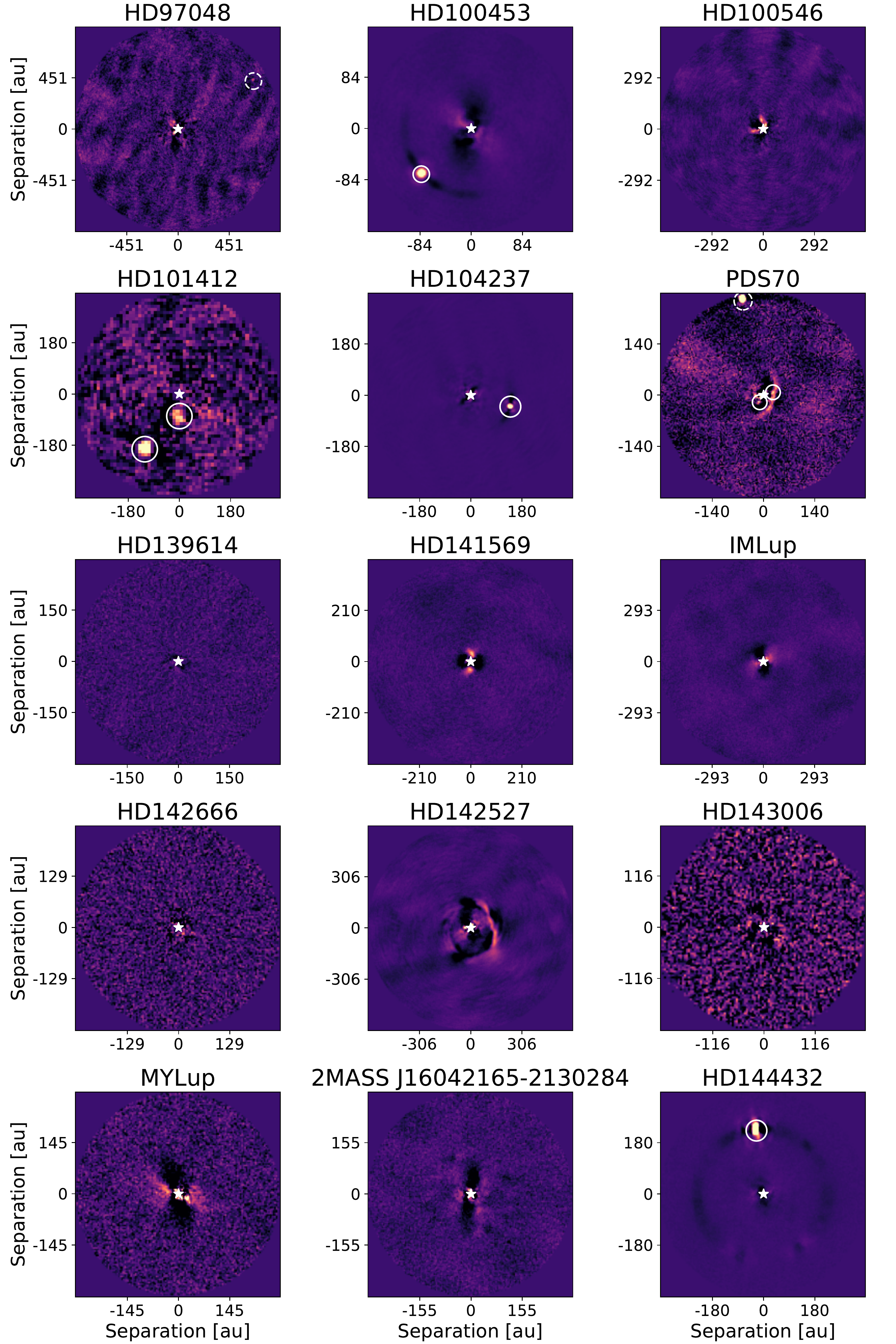}
    \caption{Final PCA-ADI residuals of the ISPY PPD sample. White solid circles indicate the position of companions, while dashed ones locate background objects.}
    \label{fig:residuals_2}
\end{figure*}

\begin{figure*}
    \centering
    \includegraphics[width=0.87\hsize]{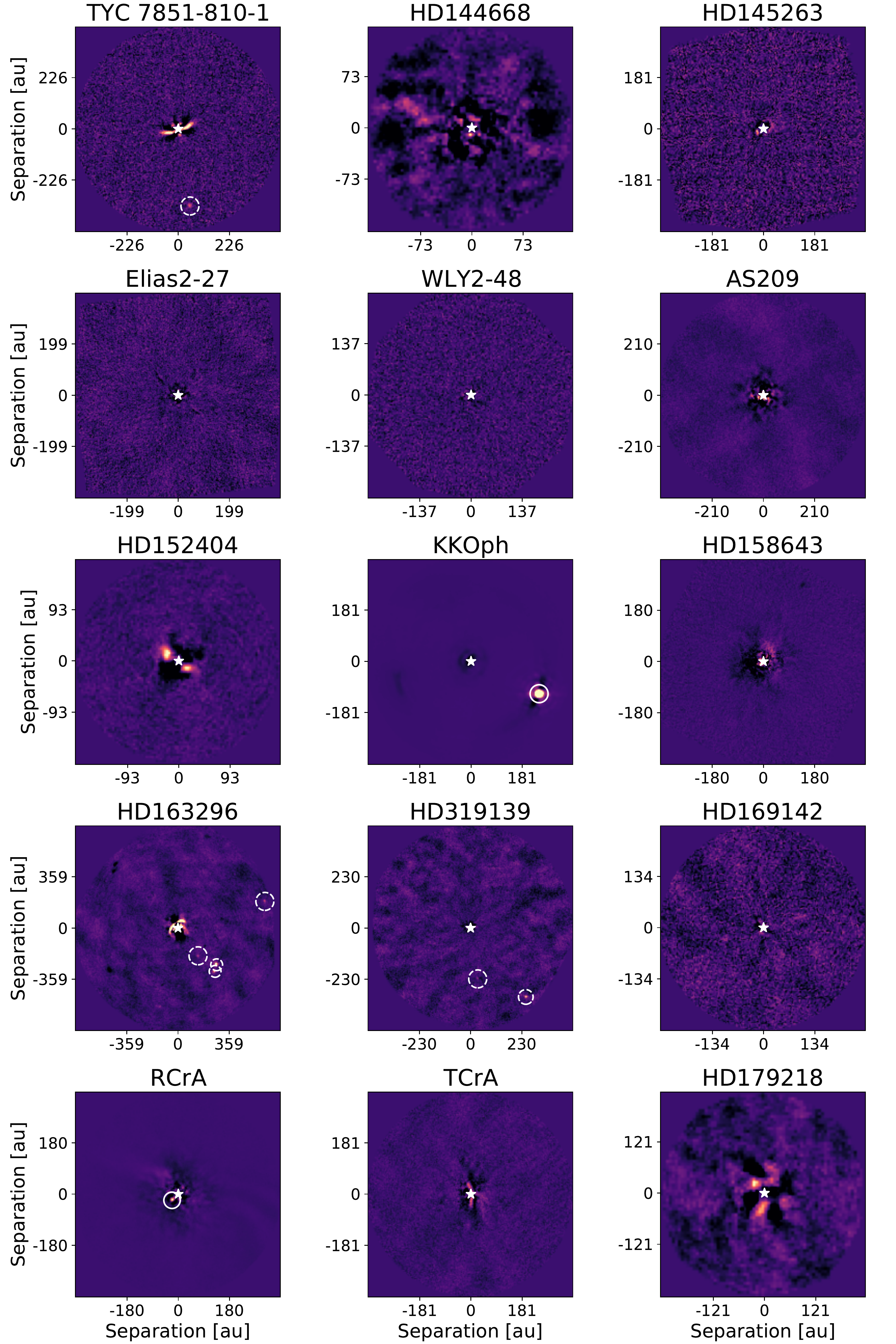}
    \caption{Final PCA-ADI residuals of the ISPY PPD sample. White solid circles indicate the position of companions, while dashed ones locate background objects.}
    \label{fig:residuals_3}
\end{figure*}

Figures~\ref{fig:residuals_1}, \ref{fig:residuals_2} and \ref{fig:residuals_3} report the residuals obtained after PSF-subtraction for the 45 targets presented in this work. Every scale is shown in au and is based on the target distance and disk extent. The color-scale is linear. Companions and companion candidates are highlighted with a white circle, while the position of indirectly inferred companions which were not detected in our images are shown with dashed white circles. White stars at the center of the images indicate the position of the primary stars.

\section{Noise properties on the residuals}
\label{app:qqplots}
\begin{figure}
    \centering
    \includegraphics[width=0.87\hsize]{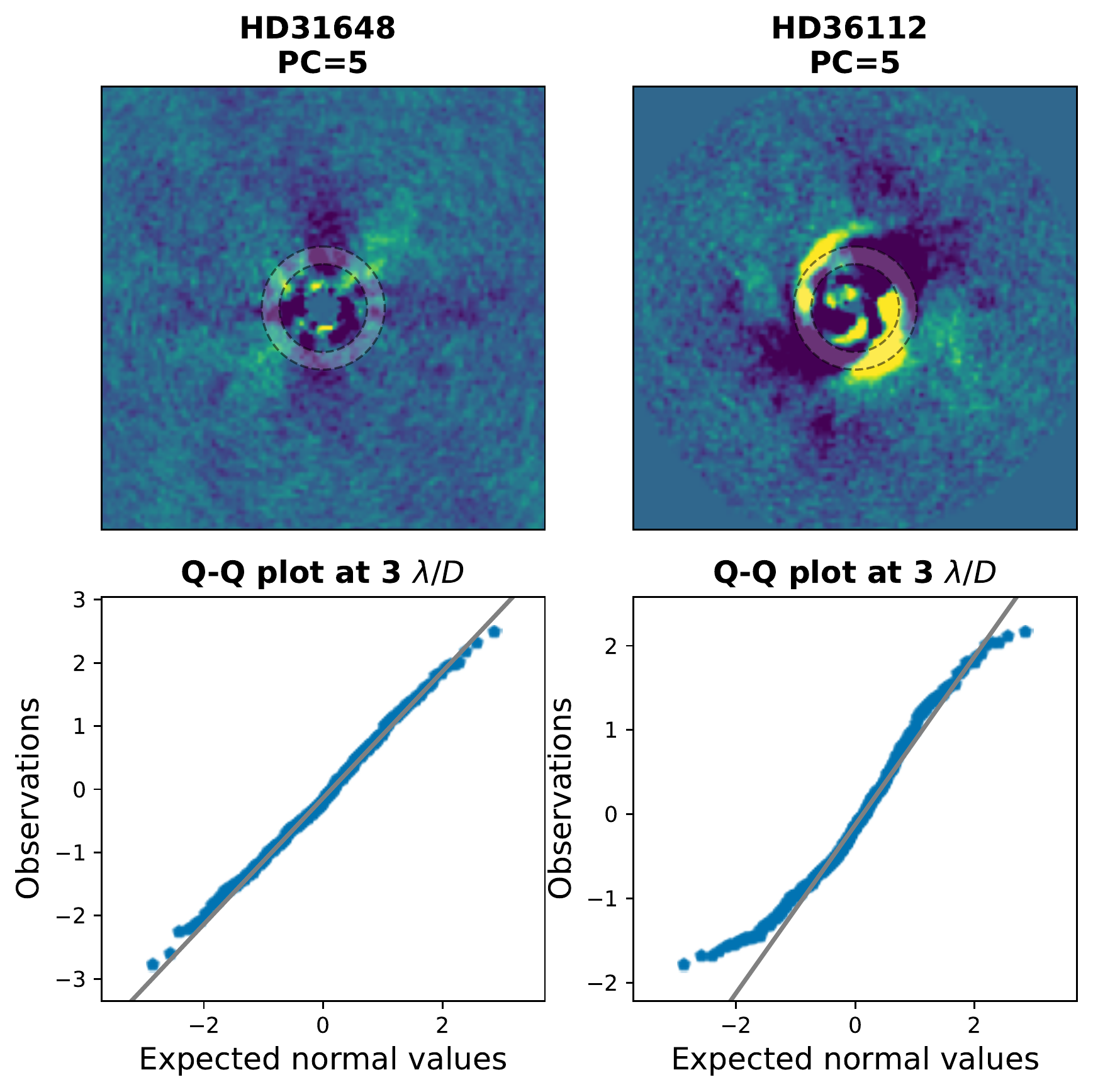}
    \caption{Noise statistics of the NaCo-ISPY data. {\it Top:} Residuals for HD31648 and HD36112 ({\it left} and {\it right} respectively). The ring highlights the location where the noise is investigated in this figure, at $3\lambda/D$. {\it Bottom:} Q-Q plots comparing the noise in the data (blue datapoints) with the expectation from Gaussian noise (gray line). In the right panel the light tails due to the presence of the disk signal and consequent negative self-subtraction cause the noise to deviate from a normal distribution.}
    \label{fig:QQplots}
\end{figure}

Here we discuss the noise properties of the NaCo-ISPY data, justifying (i) the choice for the underlying noise distribution and (ii) the exclusion from further analysis of regions of the images where bright scattered light disk signals were present. Figure~\ref{fig:QQplots} shows residuals and Q-Q plots for two of our datasets: HD31648 and HD36112. For more information of Q-Q plots we refer to \cite[submitted]{Bonse2023}. The left panel demonstrates that the noise distribution is consistent with Gaussian, even though, as explained in \cite[submitted]{Bonse2023}, we can not undisputedly prove that it is drawn from a Gaussian distribution. In the right panel we show the deviation of the light tails from normal distribution caused by the presence of strong disk signal. Furthermore, in those region the assumption of indipendent noise samples is broken given the correlation induced by the presence of an extended source spread over several resolution elements.

\section{Known detected companions}
\label{sec:known_companion}


\subsection{HD35187}
HD35187 has a known stellar companion and it appears in the Washington Double Star catalog \citep[WDS;][]{Mason2001} as WDS~05240+2458. According to the catalog, the two stars are separated by $1\farcs386\pm0\farcs005$, the PA of the companion is 192$^\circ$ and they have similar brightness ($V\approx8.5$ mag) consistent with our results reported in Table~\ref{tab:companion_candidates_results}. The spectral type of the companion is A7 \citep{Dunkin1998}. 

\subsection{HD37411}
HD37411 is a triple system first identified by \cite{Thomas2007}. The ISPY $L'$ images presented in Fig.~\ref{fig:residuals_1} reveal both companions, with their properties reported in Table~\ref{tab:companion_candidates_results}.

\subsection{NX Pup}
NX Pup (WDS J07195$-$4435AB) was revealed to be a binary system with HST images \citep{Bernacca1993} and was subsequently studied with early adaptive optics systems \citep{Brandner1995}. Our observation is consistent with early detections of the F7-G4 binary companion \citep{Schoeller1996}, and we provide astrometry and $L'$ photometry in Table~\ref{tab:companion_candidates_results}, obtained with the MCMC algorithm. 

\subsection{HD72106}
HD72106 is a binary system in which the companion is a Herbig Ae/Be star \citep{Vieira2003}. Interestingly, the primary star shows evidence of a strong magnetic field, while the secondary does not \citep{Folsom2008}. 

\subsection{HD100453}
The companion of HD100453 was first detected by \cite{Chen2006} with the VLT/NaCo instrument. Subsequent HST and NaCo observations confirmed the common proper motion with HD100453\,A \citep{Collins2009} and additional spectroscopic data presented by the same authors constrained the spectral type of the companion to be M4.0−M4.5V. \cite{Wagner2018} studied the orbit of the companion combining observations obtained with a baseline of 14 years, concluding that the companion is coplanar with the circumstellar disk surrounding HD100453\,A. 

\subsection{HD104237}
The companion detected in the ISPY data was first imaged by \cite{Grady2004}, who resolved the region around HD104237, revealing a small T association \citep{Grady2005}. Furthermore, \cite{Grady2004} demonstrated that a circumstellar disk is surrounding the companion HD104237\,B shown in Fig.~\ref{fig:residuals_2}. The other confirmed companions were at too large separations to be detected. 

\subsection{PDS70}
PDS70 harbors two forming planets within the large disk cavity \citep{Keppler2018, Muller2018, Haffert2019}. Subsequent studies tried to infer several properties from the two companions and their environment \citep[e.g.,][]{Wang2020, Wang2021, Stolker2020_pds70, Cugno2021}. The ISPY dataset presented here was first presented in \cite{Keppler2018} and then reanalyzed in \cite{Haffert2019} and \cite{Stolker2020_pds70}. For this reason, Table~\ref{tab:companion_candidates_results} reports the results obtained by \cite{Stolker2020_pds70} for PDS70\,b and those from \cite{Haffert2019} for PDS70\,c.

In addition, in the final residuals a background star north from PDS70 is visible. The nature of the object has already been assessed in \cite{Haffert2019}, and therefore we do not investigate it further. 

\subsection{HD144432}
Initially thought to be a binary system \citep[e.g.,][]{Perez2004, Carmona2007}, HD144432 was later proven to be triple system, with the B component being actually a close binary itself \citep{Muller2011}. The latter finding was obtained with NaCo imaging data at shorter wavelengths and the spectral classes of the companion binary system (K7 and M1) were inferred using VLT/FEROS spectra \citep{Muller2011}. Our $L'$ data do not allow us to distinguish the two stars B and C. Thus, we treat it as a single unresolved point source in Table.~\ref{tab:companion_candidates_results}. 

\subsection{KK Oph}
KK Oph B is a companion known since 1997 \citep{Leinert1997}. \cite{Carmona2007} analyzed VLT/FEROS2 spectra and classified the companion to have spectral type G6. 

\subsection{R\,CrA}
The young stellar companion around R\,CrA was first indirectly inferred by \cite{Takami2003}, and then simultaneously imaged with VLT/NaCo \citep{Cugno2019_b} and VLT/SPHERE \citep{Mesa2019} observations. The latter authors inferred the spectral type of the object: M3-M3.5. As the data presented here were already analyzed by \cite{Cugno2019_b}, we report their results in Table.~\ref{tab:companion_candidates_results}.

\section{Background objects}
\label{sec:bkg_obj}

\subsection{TYC 7851-810-1}

A companion candidate is seen at a separation of $2\farcs24$ from the central star. The object is also resolved in several archival SPHERE observations during 2016 and 2017 presented in \cite{Villenave2019}, but without sufficient time baseline and astrometric precision to test for common proper motion. 

This object is also resolved in the {\it Gaia} catalog, as {\it Gaia} DR3\,5997490206145064448, at a separation of $2\farcs306$ \citep{Gaia2022}. The {\it Gaia} catalog does not include parallax or proper motion for this candidate companion, so we instead study the photometry to determine whether the two objects are co-distant. The companion has contrasts of $8.61\pm0.02$~mag and $8.34\pm0.1$~mag in the $G$ and $L'$ filters respectively. The very similar $G-L'$ colors of the two objects imply similar spectral types -- this is inconsistent with the object being a true companion, given its much lower magnitude.

\subsection{HD97048}
The images of HD97048 present one companion candidate with a separation of $4\farcs27$ (northwest) from the central star. We reduced archival Gemini/NICI data in the $H+K$ band (Prog. ID:GS-2012A-C-3, PI: Honda) from 2012 in order to determine the proper motion of the companion candidate, which is consistent with a background source.

\subsection{HD163296}
Four companion candidates are identified in the residuals of HD163296. All these candidates where already investigated by \cite{Mesa2019_HD163296}, who verified them and excluded companionship.

\subsection{HD319139}
Two companion candidates are identified in the residuals of HD319139. 
Comparing the images with archival HST images (Prog ID:10348, observation date: 2005-08-03, PI: Herczeg) we could verify that they are all background objects not comoving with the star.

\section{Detection probability maps comparison}

\begin{figure*}[t!]
    \includegraphics[width=\hsize]{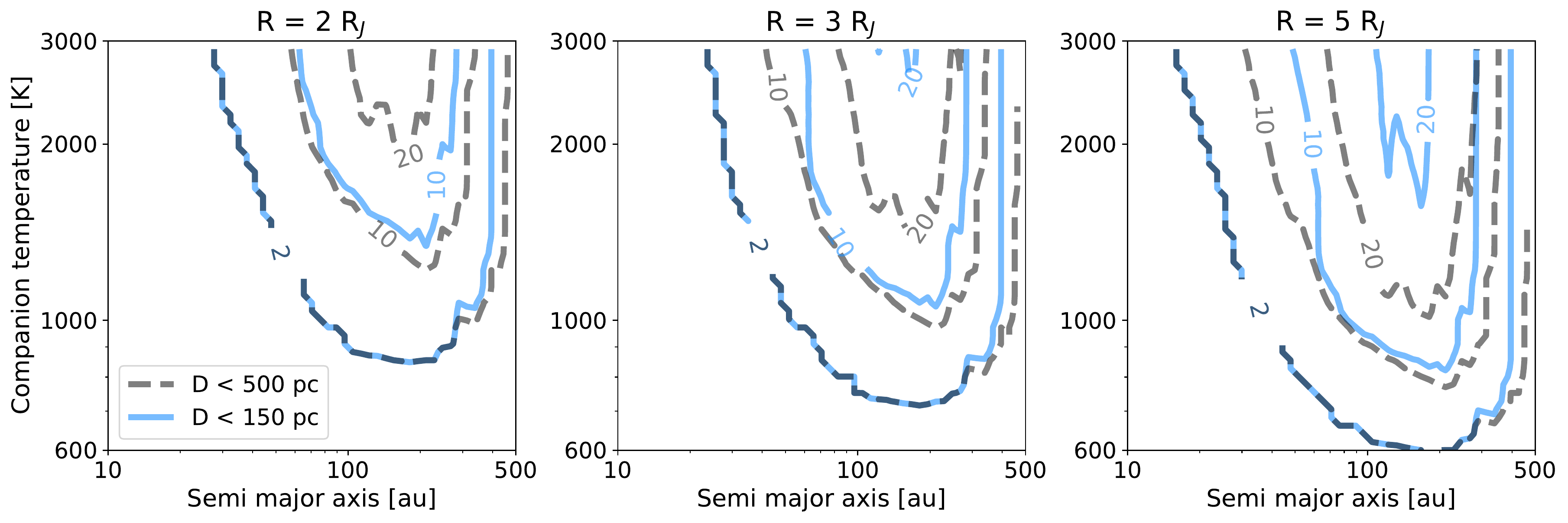}
    \caption{Detection probability maps obtained from the subsample with $d<150$~pc. The three plots represent maps when assuming $\Rp=2,3,5~\RJ$ ({\it left, middle} and {\it right} panels, respectively). Solid contours limit the areas with completeness $2$, $10$ and $20$, while dotted lines are reported from Fig.~\ref{fig:prob_map} and limit the same completeness values when the entire sample of 45 targets is considered, thus allowing the direct comparison between the two cases and highlighting the dominance of nearby targets when searching for protoplanets and interpreting statistical results.}
    \label{fig:map_distance}
\end{figure*}

\begin{figure*}[t!]
    \includegraphics[width=\hsize]{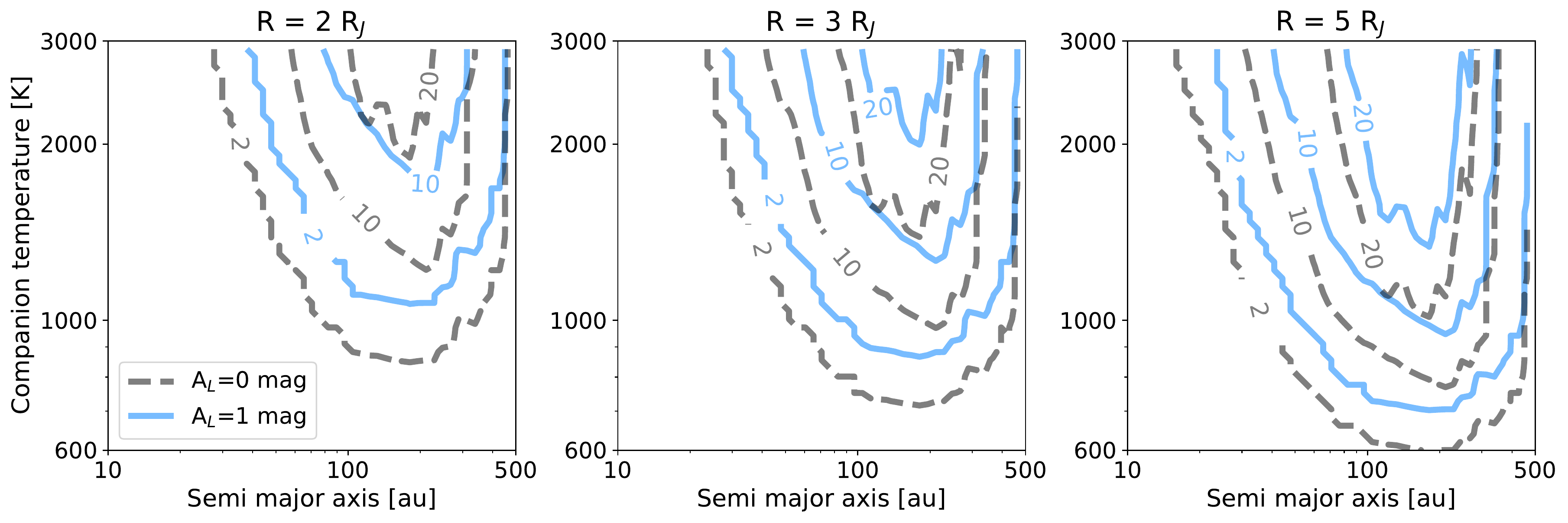}
    \caption{Survey completeness maps obtained considering an additional extinction factor $A_{L'}=1.0$~mag. The three plots represent maps when assuming $\Rp=2,3,5~\RJ$ ({\it left, middle} and {\it right} panels, respectively). Solid contours limit the areas with completeness $2$, $10$ and $20$, while dotted lines are reported from Fig.~\ref{fig:prob_map} and limit the same detection probability values when no extinction effects are included, thus allowing the direct comparison between the two cases.}
    \label{fig:map_extinction}
\end{figure*}

Here we present the comparison of detection probability maps to low-mass companions when (i) considering only a fraction of the entire sample, namely targets with $d<150$~pc, and (ii) when roughly including some extinction from circumstellar disk material potentially attenuating the planetary flux able to escape the circumstellar disk. Case (i) is presented in Fig.~\ref{fig:map_distance}, where only the 23 targets are included when generating the survey map. Similarly, case (ii) is shown in Fig.~\ref{fig:map_extinction}, where we used contrast curves 1.0 mag brighter at each separation due to the extinction from potential circumstellar and circumplanetary disk material.

\end{appendix}

\end{document}